\documentclass[12pt]{article}
\usepackage{amsmath}
\usepackage{amssymb}
\usepackage{amsthm}
\usepackage{graphicx}
\usepackage{lineno}
\usepackage{float}
\usepackage[all]{xy}
\usepackage{mathrsfs}

\usepackage{xcolor}
\usepackage{tikz}
\usetikzlibrary{tqft}

\newtheorem{theorem}{Theorem}

\newcommand{\tr}{{\rm tr}}

\newcommand{\A}{\mathcal{A}}
\newcommand{\B}{\mathcal B}

\begin{document}

\title{\bf The physical meaning of the holographic principle}

\author{{Chris Fields$^a$, James F. Glazebrook$^{b,c}$ and Antonino Marcian\`{o}$^{d,e,f}$}\\ \\
{\it$^a$ Allen Discovery Center, Tufts University, Medford, MA 02155 USA}\\
{fieldsres@gmail.com}\\
{ORCID: 0000-0002-4812-0744}\\
{\it$^b$ Department of Mathematics and Computer Science,} \\
{\it Eastern Illinois University, Charleston, IL 61920 USA} \\
{\it$^c$ Adjunct Faculty, Department of Mathematics,}\\
{\it University of Illinois at Urbana-Champaign, Urbana, IL 61801 USA}\\
{jfglazebrook@eiu.edu}\\
{ORCID: 0000-0001-8335-221X}\\
{\it$^d$ Center for Field Theory and Particle Physics \& Department of Physics} \\
{\it Fudan University, Shanghai, CHINA} \\
{marciano@fudan.edu.cn} \\
{\it$^e$ Laboratori Nazionali di Frascati INFN, Frascati (Rome), Italy, EU} \\
{marciano@lnf.infn.it} \\
{\it$^f$ INFN sezione Roma ``Tor Vergata'', 00133 Rome, Italy, EU} \\
{ORCID: 0000-0003-4719-110X}
}

\maketitle

{\bf Abstract} \\
We show in this pedagogical review that far from being ``an apparent law of physics that stands by itself'' (R. Bousso, Rev. Mod. Phys. 74 (2002), 825--874), the holographic principle (HP) is a straightforward consequence of the quantum information theory of separable systems.  It provides a basis for the theories of measurement, time, and scattering.  Principles equivalent to the HP appear in both computer science and the life sciences, suggesting that the HP is not just a fundamental principle of physics, but of all of science.

\tableofcontents

\pagebreak

\begin{flushright}
The world is all that is the case. \\
-- L. Wittgenstein \\
~ \\
No question?  No answer! \\
-- J. A. Wheeler \\
\end{flushright}

\section{Introduction}

The Holographic Principle (HP) was originally stated as a conjecture by `t~Hooft \cite{tHooft:93}:

\begin{quote}
Given any closed surface, we can represent all that happens inside it by degrees of freedom on this surface itself.
\end{quote}

\noindent
The HP generalizes Bekenstein's area law \cite{bekenstein:73} for the entropy of a black hole (BH):
\begin{equation}
\label{Bek}
S = \frac{A}{4},
\end{equation}
where $S$ denotes the thermodynamic entropy of a BH and $A$ its horizon area in Planck units.  The number of ``degrees of freedom on this surface itself'' cannot, in particular, exceed $\frac{A}{4}$, with $A$ the area of the surface in question \cite{tHooft:93}.

Susskind \cite{susskind:95} provided the first physical implementation of `t Hooft's conjecture by defining an explicit mapping from volume to surface degrees of freedom for a general closed system. This mapping assumes that all light rays that are normal to any element of surface within the bulk are also normal to the boundary.  Bousso \cite{bousso:02} then showed that requiring covariance induces a holographic limit on information transfer by light, reformulating Eq. \eqref{Bek} as a covariant entropy bound:
\begin{equation}
\label{Bo}
S(L(\Sigma)) \leq ~\frac{A(\Sigma)}{4},
\end{equation}
where $A(\Sigma)$ denotes the area in Planck units of a (typically but not necessarily) closed surface $\Sigma$, and $L(\Sigma)$ any light-sheet of $\Sigma$, defined as any collection of converging light rays that propagate from $\Sigma$ toward some focal point away from $\Sigma$.  Bekenstein's area law emerges as the special case in which the equality in \eqref{Bo} holds.  Bousso also provided several counterexamples showing the failure of a straightforward interpretation of the HP as a spacelike entropy bound.

When formalized by \eqref{Bek} or \eqref{Bo}, the HP is semiclassical; indeed it is ``quantum'' only in its reliance on Planck units and hence a finite value of $\hbar$.  The entropy $S$ is a classical thermodynamic entropy.  In the context of general relativity (GR), $\Sigma$ is a continuous classical manifold enclosing a continuous classical volume characterized by a real-valued metric.  As `t~Hooft \cite{tHooft:93} pointed out, the HP renders $S(L(\Sigma))$ independent of the metric inside $\Sigma$:

\begin{quote}
The inside metric could be so much curved that an entire universe could be squeezed inside our closed surface, regardless how small it is.  Now we see that this possibility will not add to the number of allowed states at all.
\end{quote}
\noindent
Here the ``allowed states'' are the thermodynamic states of $\Sigma$, states that an external observer can count by measuring energy transfer between the system and its external environment.  In the case of BH, Rovelli \cite{rovelli:17, rovelli:19} has shown explicitly that states that are effectively isolated from the external environment and hence do not contribute the system--environment interaction over relevant time scales do not contribute to $S(L(\Sigma))$.  A BH can, in principle, have arbitrarily many such isolated states, as in Wheeler's ``bag of gold'' model of a BH with a small horizon and large interior \cite{almheiri:21}.

The HP was given broader theoretical relevance within quantum gravity (QG) research by Maldacena \cite{maldecena:98}, who showed that a string QG on a ``bulk'' $d$-dimensional anti de Sitter (AdS) spacetime and a conformal quantum field theory (CFT) on its $d-1$-dimensional boundary encode the same information.  A more limited dS/CFT holographic duality has also been explored \cite{strominger:01}.  While such dualities have seen wide theoretical application, their physical motivation remains that of `t~Hooft's conjecture and Bekenstein's area law. The existence of these holographic dualities suggests that the HP is both deep and fully general, but they do not explain why this should be the case.  Bousso \cite{bousso:02} summarized the situation by remarking that the HP remains:

\begin{quote}
$\dots$ an apparent law of physics that stands by itself, both uncontradicted and unexplained by existing theories, that may still prove incorrect or merely accidental, signifying no deeper origin.
\end{quote}

Our purpose here is to respond to this remark of Bousso's by developing a clear, consistent, and at bottom, a very simple picture of the physical meaning of the HP.  Building on previous results \cite{fm:19, fm:20, fg:20, fgm:21, addazi:21, fgm:22}, we show that the HP can be generalized to describe the maximum classical information flow implemented by any physical interaction between mutually separable, i.e. unentangled, finite physical systems.  Specifically, we can state:

\begin{quote}
{\em Generalized Holographic Principle} (GHP): If $U = AB$ is a finite closed system, then if $|AB \rangle$ is separable, the classical information exchange between $A$ and $B$ is limited to $N$ bits, where $N$ is the dimension of the interaction Hamiltonian $H_{AB}$.
\end{quote}

This GHP is a fully quantum information-theoretic principle that is entirely independent of geometric considerations.  The $N$ exchanged bits can, however, without loss of generality be viewed as encoded at a density of no more than 1 bit per 4 $l_P^2$ on an ancillary, spacelike boundary $\mathscr{B}$ separating $A$ from $B$.  Bousso's covariant formulation follows as a special case whenever $\mathscr{B}$ is considered a physical boundary traversed by a light sheet, i.e. whenever photons (or indeed any gauge bosons) are considered to be ``carriers'' of the exchanged information.

After setting up the formalism in \S 2, we show in subsequent sections how:

\begin{enumerate}
\item The GHP enables a provably general theory of quantum measurement \cite{ffgl:22} that is fully consistent with the free-energy principle (FEP) introduced by Friston and colleagues \cite{friston:10, friston:13, ramstead:18, friston:19, ramstead:22} as a description of active inference by Bayesian agents (\S 3).
\item The GHP provides a natural definition of system-relative entropic time applicable to any physical system $A$ that is separable from its environment $B$ (\S 4).
\item The GHP allows us to view $\mathscr{B}$ as a scattering center and its areal elements as encoding S-matrix elements (\S 5).
\end{enumerate}

When generalized to the GHP, therefore, the HP does not ``stand by itself'' but is rather a fundamental principle from which much familiar physics follows.  It has, moreover, a simple and intuitively obvious physical meaning:

\begin{quote}
{\em GHP} (informal): Any classical information exchanged between finite physical systems is encoded on the boundary between them.
\end{quote}

Intersystem boundaries are, in other words, classical information channels.  We review in \S 6 various statements of this same principle that have been derived in statistical physics, computer science, and the life sciences.  We conclude that the HP is a foundational principle not just of physics, but of all of science.

\section{Holographic screens are information encoding boundaries} \label{screens}

\subsection{Physical interaction is information exchange} \label{exchange}

Let $U$ be an isolated, finite-dimensional quantum system and consider an arbitrarily-chosen bipartite decomposition $U = AB$ corresponding to a Hilbert-space tensor product $\mathcal{H}_U = \mathcal{H}_A \otimes \mathcal{H}_B$.  Any such decomposition induces an interaction $H_{AB} = H_U - (H_A + H_B)$.  Consider a time period that is short enough that $H_{AB}$ is effectively constant, and during which $H_{AB}$ is weak enough that $|AB(t) \rangle$ can be considered separable, i.e. $|AB(t) \rangle = |A(t) \rangle |B(t) \rangle$ over the entire time period of interest.  In this case, we can choose orthogonal basis vectors $|i^k \rangle$ so that:

\begin{equation} \label{ham}
H_{AB} = \beta_k K_B\, T_k \sum_i^N \alpha^k_i M^k_i,
\end{equation}
\noindent
where $K_B$ denotes Boltzmann's constant, $T$ is the absolute temperature of the environment, $k =~A$ or $B$, the $M^k_i$ are $N$ mutually-orthogonal Hermitian operators with eigenvalues in $\{ -1,1 \}$, the $\alpha^k_i \in [0,1]$ are such that $\sum^N_i \alpha^k_i = 1$, and $\beta_k \geq \ln 2$ is an inverse measure of $k$'s thermodynamic efficiency that depends on the internal dynamics $H_k$; see \cite{fm:19, fm:20, fg:20, fgm:21, addazi:21} for further motivation and details of this construction.  For fixed $k$, the operators $M^k_i$ clearly must commute, i.e. $[M^k_i, M^k_j] = M^k_i M^k_j - M^k_j M^k_i = 0$ for all $i, j$; hence when expressed as Eq. \eqref{ham}, $H_{AB}$ is swap-symmetric under the permutation group $S_N$ for each $k$.  We can, therefore, write $N = \dim(H_{AB})$, i.e. the eigenvalues $H_{AB}$ can be encoded by $2^N$ distinct $N$-bit strings.  The weak-interaction limit requires $N \ll \dim(\mathcal{H}_A), \dim(\mathcal{H}_B)$, although as discussed below, this condition is not sufficient to guarantee separability.

\begin{figure}[H]
\centering
\includegraphics[width=13 cm]{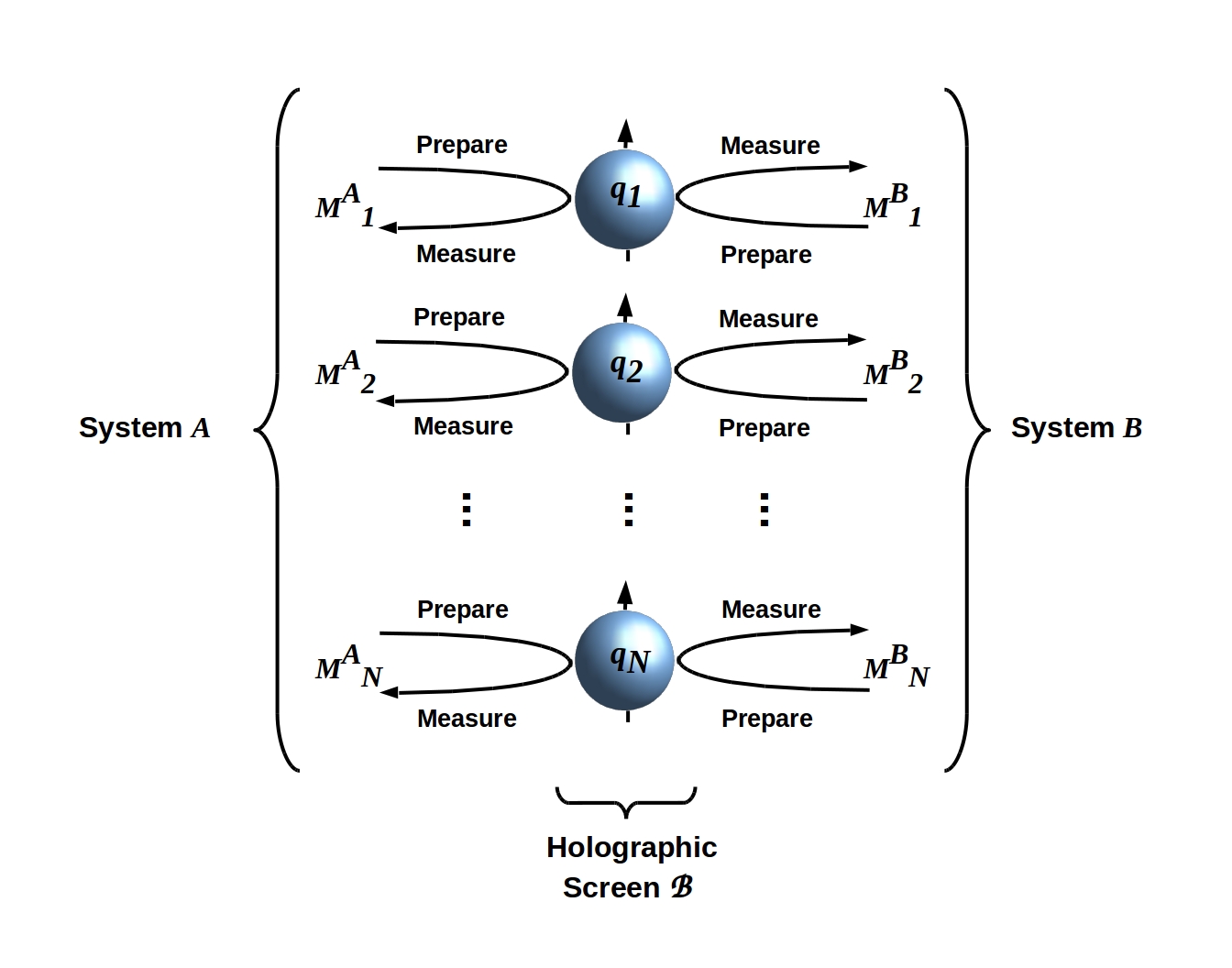}
\caption{A holographic screen $\mathscr{B}$ separating finite systems $A$ and $B$ with an interaction $H_{AB}$ given by Eq. \eqref{ham} can be realized by an ancillary array of noninteracting qubits that are alternately prepared by $A$ ($B$) and then measured by $B$ ($A$).  Qubits are depicted as Bloch spheres \cite{nielsen:00}.  There is no requirement that $A$ and $B$ share preparation and measurement bases, i.e. QRFs.  Adapted from \cite{fgm:21} Fig. 1, CC-BY license.}
\label{qubit-screen-fig}
\end{figure}

When expressed as Eq. \eqref{ham}, $H_{AB}$ can be realized physically as illustrated in Fig. \ref{qubit-screen-fig}.  The operators $M^k_i$ are interpreted, as the notation suggests, as measurement operators, and dually as state-preparation operators \cite{pegg:02}.  As each of the $M^k_i$ has eigenvalues in $\{ -1,1 \}$, they can be regarded as $z$-spin operators $s^k_{z(i)}$ acting on individual qubits $q_i$.  The orthogonality of the $M^k_i$ requires the $q_i$ to be mutually independent, i.e. non-interacting.  Each ``cycle'' of interaction between $A$ and $B$ then comprises four sequential steps: preparation of the $q_i$ by $B$, measurement of the $q_i$ by $A$, preparation of the $q_i$ by $A$, and measurement of the $q_i$ by $B$.  The systems $A$ and $B$ thus exchange $N$ bits of classical information on each cycle.  Note that, in this picture, the operators $M^A_i$ and $M^B_i$ do not act directly on the Hilbert spaces $\mathcal{H}_B$ and $\mathcal{H}_A$ of $B$ and $A$ respectively, but on the $N$-dimensional effective Hilbert spaces $\mathcal{H}^A_{<q_i>}$ and $\mathcal{H}^B_{<q_i>}$ that specify, from the perspectives of $A$ and $B$ respectively, the states of the $q_i$.  As the eigenvalues of the $M^k_i$ when considered together encode an eigenvalue of $H_{AB}$, the classical information exchanged is the current eigenvalue of $H_{AB}$, i.e. the energy transferred by the interaction.  The symmetry of the interaction cycle then assures conservation of energy.

The array $q_i$ of noninteracting qubits via which $A$ and $B$ exchange classical information clearly performs the functions of a holographic screen:

\begin{itemize}
\item The $q_i$ separate $A$ from $B$.  The interpretation of the $M^k_i$ as preparation and measurement operators depends critically on the assumption of separability; if $A$ and $B$ are entangled, i.e. if $|AB \rangle$ fails to factor as $|AB \rangle = |A \rangle |B \rangle$, the idea that $A$ ``prepares'' or ``measures'' $|B \rangle$ is physically meaningless.
\item The $q_i$ encode all of the classical information about $B$ accessible to $A$ and vice versa.  Indeed if $H_U$ remains unspecified, $H_A$ and $H_B$ can vary arbitrarily without affecting $H_{AB}$, as required by `t~Hooft's idea of squeezing ``an entire universe'' into $B$ without affecting $A$.
\end{itemize}

\begin{figure}[H]
\centering
\includegraphics[width=10 cm]{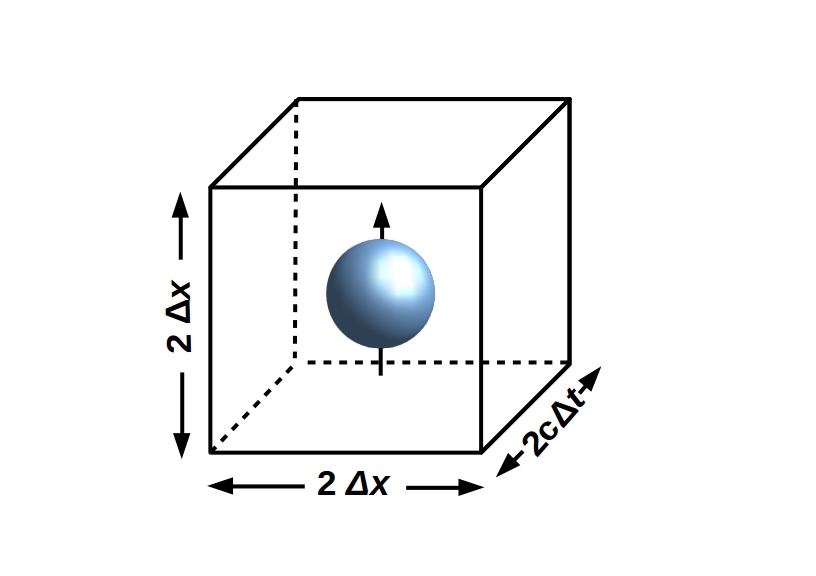}
\caption{One qubit degree of freedom (represented as a Bloch sphere), e.g. a spin, embedded in a 3d voxel at some minimal scale $\Delta x$, $\Delta t$.  Here $c$ is the minimum speed of (classical) information transfer.}
\label{qubit-fig}
\end{figure}

We can, therefore, think of the $q_i$ as ``points'' or more accurately ``sites'' on a boundary $\mathscr{B}$ separating $A$ from $B$ and hence separating their respective Hilbert spaces $\mathcal{H}_A$ and $\mathcal{H}_B$, where clearly $AB = U$ requires $\mathcal{H}_A \otimes \mathcal{H}_B = \mathcal{H}_U$.  This boundary $\mathscr{B}$ is, however, entirely ancillary; its states $|q_i \rangle$ are not elements of $\mathcal{H}_U$.  The independence of the $q_i$ gives $\mathscr{B}$ a discrete, indeed, a Grothendieck topology (see e.g. \cite{macclane:94}).  At this stage of the construction, $\mathscr{B}$ is just a discrete topological space that is neither characterized as a space-like nor as a time-like surface. The embedding of $\mathscr{B}$ in a $d+1$ dimensional spacetime manifold can be achieved through a tessellation in voxels of the latter. Voxels that neighbour each other and cubulate spacetime introduce a concept of distance between qubits. In turn qubits define the nodes of a one-complex that represent the discretization of $\mathscr{B}$. We can provide an illustrative example by embedding $\mathscr{B}$ in a $2+1$ dimensional spacetime manifold. In this case, we can ``geometrize'' $\mathscr{B}$ as shown in Fig. \ref{qubit-fig} by embedding each of the $q_i$ in a (conventionally $\emph{3d}$) voxel of size $(2\Delta x)^2 \cdot 2c\Delta t$ where to preserve covariance and hence Eq. \eqref{Bo}, $\Delta x \geq l_P$ and $\Delta t \geq t_P$, with $l_P$ and $t_P$ denoting the Planck length and time respectively, and $c$ is the maximal speed of classical information transfer, i.e. the speed of light.  As $\mathscr{B}$ itself is ancillary to $\mathcal{H}_U$, this geometry is ancillary to $\mathcal{H}_U$.  The geometry on $\mathscr{B}$ has, therefore, no effect on the physics implemented by the joint system self-interaction $H_U$ or the $A-B$ interaction $H_{AB}$.

The GHP as stated above clearly follows immediately, by construction, from the mutual separability of $A$ and $B$ via Eq. \eqref{ham}; Eq. \eqref{Bo} and hence the covariant HP follow when the boundary $\mathscr{B}$ is geometrized as above.  The GHP can, therefore, be viewed as an alternative way of stating the fundamental idea that physical systems must have their own, mutually conditionally independent states if they are to be regarded as interacting.  This idea of conditional independence is so deeply embedded in our language -- the language of ``things'' that interact with each other -- that it is seldom made explicit.  The GHP formalizes an obvious logical consequence of this idea: finite things can only exchange finite information by interacting, and this information has to fit through the finite channel implemented by the boundary that separates them.  We will see in what follows that this seemingly-simple fact has significant implications both in physics and in other disciplines.  Indeed, it generates, with minimal further assumptions, much of what is considered foundational in physics, computer science, and the life sciences.

\subsection{Information is strictly conserved} \label{conservation}

The representation of $H_{AB}$ as a Hermitian operator in Eq. \eqref{ham} depends on the Axiom of Unitarity \cite{vonN:55}; in particular, it requires the time evolution operator:

\begin{equation}
\mathcal{P}_U = ~e^{-(\imath / \hbar)H_U (t)}
\end{equation}
\noindent
to be unitary.  The Axiom of Unitarity guarantees that time evolution is reversible, and hence that information is conserved, in any closed system.

Because the GHP follows by construction from Eq. \eqref{ham}, it is clearly consistent with the Axiom of Unitarity, and hence with strict conservation of information.  Indeed, the informational symmetry of any holographic screen $\mathscr{B}$ enforces the conservation of information by preventing any ``build up'' of information on one side or the other.

If we regard the conservation of information (and hence unitarity) as a fundamental principle analogous to the conservation of energy, then we can formulate it as the Principle that the net information in any closed system remains constant.  The joint system $U$ is closed by definition; hence unitarity requires the net information content of $U$ to be constant.  We can, therefore, rescale the net information in $U$ to zero.  This applies, in particular, to net classical information:

\begin{quote}
Conservation of Classical Information (CCI): The net classical information in any closed system is zero.
\end{quote}

If we consider {\em irreversibly encoded} classical information, to which Landauer's Principle \cite{landauer:61, landauer:99, bennett:82} applies, CCI clearly follows from the conservation of energy: if the net energy of a system is zero, the net irreversibly encoded classical information in that system can only be zero.  Compliance with CCI is guaranteed if we require:

\begin{quote}
Exclusive Holographic Encoding (EHE): Classical information is encoded only on holographic screens.
\end{quote}

Indeed the informational symmetry of holographic encoding renders EHE and CCI equivalent.  Both can be viewed as ``no collapse'' principles that render classical information strictly ancillary to the closed-system dynamics $H_U$.  Classical information encoded on $\mathscr{B}$ is not, however, ancillary to either of the separated systems $A$ or $B$; this encoded information is input to, or dually output from, $A$ or $B$ by Eq. \eqref{ham}.  From the perspective of either $A$ or $B$, $\mathscr{B}$ encodes $N \neq ~0$ bits of classical information whenever $H_{AB} \neq ~0$.  This is true, for some positive value of $N$, for any tensor-product decomposition of any closed system $U$ that meets the separability criteria that allow writing Eq. \eqref{ham}.  Hence we can restate EHE as:

\begin{quote}
Relativity of Classical Information (RCI): All classical information is decomposition-relative.
\end{quote}

We will see below that RCI renders both all observable ``systems'' and all classical memory observer-relative.  It thus generalizes the idea --- that appears in relative-state \cite{everett:57, tegmark:98}, relational \cite{rovelli:96}, and QBist \cite{fuchs:10, mermin:18} approaches to quantum theory --- that quantum {\em states} are observer-relative to quantum systems and, critically, to all classically-encoded records of previous observations.  It thus renders all such systems and records observer-dependent and hence ``non-objective.''  There is, however, from a theoretical perspective nothing controversial about RCI; it merely restates the Axiom of Unitarity.  We suggest, as Bohr \cite{bohr:34} and Mermin \cite{mermin:98} have before us, that ``what quantum theory is trying to tell us'' is precisely RCI.

\subsection{Classical encoding requires free choice of basis} \label{choice}

Writing Eq. \eqref{ham} requires choosing the basis vectors $|i^k \rangle$, where again $k = ~A$ or $B$.  In the physical realization of $\mathscr{B}$ as a qubit array shown in Fig. \ref{qubit-screen-fig}, choosing the $|i^k \rangle$ is choosing, for each of $A$ and $B$, the $z$ axis used to measure $s_z$ for each of the $q_i$.  This choice determines, for each of $A$ and $B$, which of the $2^N$ eigenvalues of $H_{AB}$ is encoded on $\mathscr{B}$.  Choosing the $|i^k \rangle$ is, therefore, effectively choosing the zero-point of energy for each of $A$ and $B$.  These zero points can clearly be different.

Free choice of the $|i^k \rangle$ by us as theorists is equivalent, operationally, to free choice of $|i^A \rangle$ and $|i^B \rangle$ by $A$ and $B$, respectively.  ``Free choice'' is standardly interpreted as freedom from local determinism, e.g. by events in the past light cone \cite{conway:09}; while unitarity of closed-system (e.g. $U$) evolution can be read as a form of superdeterminism \cite{tipler:14}, this is fully consistent with local free choice by open systems such as $A$ or $B$.  In the present context, free choice of $|i^A \rangle$ and $|i^B \rangle$ by $A$ and $B$, respectively, requires that neither $|i^A \rangle$ nor $|i^B \rangle$ is fixed by the classical data encoded on $\mathscr{B}$, the only data available to either $A$ or $B$.  As choosing the basis is prerequisite for assigning values to these data, this condition clearly holds.

It is useful, however, to consider the conditions under which free choice of basis fails for either $A$ or $B$.  Suppose $B$'s choice of $|i^B \rangle$ determines $|i^A \rangle$.  As the classical data encoded on $\mathscr{B}$ cannot determine $|i^A \rangle$, this situation can occur only if $A$'s choice of basis $|i^A \rangle$ depends on the state $|B \rangle$, i.e. it can occur only if $A$ and $B$ share quantum information.  In this case, however, the joint state $|AB \rangle$ is no longer separable and Eq. \eqref{ham} no longer holds.  Free choice of basis is, therefore, required for separability and hence for the GHP.  We will see below that this has significant consequences for the theory of measurement.

\section{The GHP enables a fully-general quantum theory of measurement} \label{theory-meas}

\subsection{Measurement produces finite-resolution, classical outcomes} \label{outcomes}

Quantum theory is traditionally considered to have a ``measurement problem''; indeed since Schr\"{o}dinger first introduced his cat \cite{schrodinger:35}, an enormous literature has devoted itself to the question of how observers can obtain classical information from a quantum world (see \cite{landsman:07} for a thorough review and \cite{cabello:15} for a recent compendium of philosophical positions).  We will, in this section, employ quantum theory and the GHP to construct a fully-general theory of measurement; this construction was developed in \cite{fm:19, fm:20, fg:20, fgm:21, addazi:21, fgm:22}, to which readers are referred for further details.  We show in \cite{ffgl:22} that this theory reduces, in its classical limit, to the well-established theory of active inference derived from the classical FEP \cite{friston:10, friston:13, ramstead:18, friston:19, ramstead:22}.  We suggest that this holographic quantum theory of measurement obviates the traditional measurement problem, though aside from the brief comments made earlier, we defer discussion of its relations to any of the plethora of philosophical interpretations of quantum measurement to future work.

The goal of measurement is to obtain recordable, reportable observational outcomes that can be compared to the outcomes of measurements carried out at other times or by other observers.  Such measurement outcomes must be encodable in a thermodynamically irreversible way as classical information on classical memory devices, e.g. pieces of paper, transistor arrays, or weight values on connections in a neural network.  As thermodynamically irreversible classical encoding has a finite energy cost of at least ln 2 $k_B T$ per bit \cite{landauer:61, landauer:99, bennett:82}, any such observational outcome must be encodable as a finite bit string.  This is in fact obvious -- no physical apparatus has infinite resolution -- but is often neglected when observational outcomes are represented by unrestricted real numbers.

A theory of measurement must, therefore, address three questions:

\begin{itemize}
\item It must provide a formal mechanism for mapping physical interactions to finite classical encodings of observational outcomes.
\item It must provide a formal mechanism -- operationally, a semantics -- that enables encoded observational outcomes to be meaningfully compared.
\item It must provide a mechanism that supplies the free energy required to support irreversible classical encoding.
\end{itemize}

A theory of measurement must, in other words, enable saying what it means, both operationally and thermodynamically, to claim to have measured a length of 0.300 $\pm$ 0.001 m by interacting with a wooden board via a meter stick, and then to have written down the result on a piece of paper.

The GHP enables addressing these questions precisely by localizing available classical data, available recordable memory, and available free energy to the single boundary $\mathscr{B}$.  This co-localization of informational and thermodynamic resources to $\mathscr{B}$ has three immediate consequences:

\begin{itemize}
\item A proper subset or sector $F$ of the bits encoded on $\mathscr{B}$ are accessible to an ``observer'' $A$ only as free energy that can be employed to fund processing other inputs or writing data to memory.
\item Any observational outcomes recorded by $A$ must be written on $\mathscr{B}$, and are therefore exposed to the ``world'' $B$.
\item Obtaining new observational outcomes and recording previous ones compete for free energy resources, with the measurement resolution and hence allocated numbers of bits as the tradeoff parameter.
\end{itemize}

The GHP requires, in other words, that ``observers'' be treated as physical systems subject to resource constraints.  The symmetry between state preparation and measurement required by Eq. \eqref{ham}, moreover, renders $\mathscr{B}$ informationally symmetric: $A$ and $B$ both have access to exactly the same $N$ qubits.  The labels ``observer'' and ``world'' are, therefore, for convenience only; the two parties to the interaction have exactly the same roles as physical systems.  This has a further important consequence: the classical idea of ``passive observation'' is ruled out in principle.  Obtaining information from $B$ requires acting irreversibly on $B$, expending free energy in the process.  As Wheeler \cite{wheeler:89} put it, ``No question? No answer!''

\subsection{Measurements are given meaning by quantum reference frames} \label{QRFs}

Observational outcomes are rendered comparable, and hence physically meaningful, through the use of reference frames (RFs).  Outcomes of length measurements, for example, are rendered comparable and hence meaningful by being assigned units, meters, that refer to the standardized definition of a meter, and via this to the intrinsically-spatial concept of the speed of light.  Operationally, any RF attaches units of measurement, and hence a semantics, to an observational outcome.  Outside of metrology or the laboratory, the physical implementations of RFs are often neglected in classical physics.  When considered in the context of quantum theory, any RF must be physically implemented by a quantum system and therefore must be considered a quantum RF (QRF) \cite{aharonov:84, bartlett:07}.  Meter sticks, clocks, even the Earth's gravitational and magnetic fields are physically-implemented RFs and hence are QRFs, as are all items of laboratory apparatus.

Let $Q$ be a QRF with an internal dynamics $H_Q$.  As emphasized by \cite{bartlett:07}, by virtue of being a quantum system $Q$ encodes ``nonfungible'' information, i.e. information that cannot be written as a finite bit string.  This nonfungible information can be thought of as the quantum phase information encoded by an instantaneous pure state $|Q \rangle$.  The existence of this nonfungible information indeed follows from the GHP: the information about $Q$ obtainable by any external observer $A$ is limited to the eigenvalue of $H_{QA}$ finitely encoded on their mutual boundary.  This finite encoding does not, in principle, fully specify $H_Q$.

Operationally, any QRF $Q$ is a physical implementation of a quantum computation:

\begin{equation}
Q: \{ 0,1 \}^n \rightleftarrows \{ 0,1 \}^m
\end{equation}

that reversibly maps between ``raw'' data representable as $n$-bit strings and meaningful observational outcomes representable as $m$-bit strings, with the forward mapping implementing measurement and the reverse mapping implementing preparation.  As developed in previous work \cite{fg:20, fgm:21} and proven in the general case in \cite{fgm:22}, any such computation can be given a category-theoretic representation as a {\em Cone-Cocone Diagram} (CCCD):

\begin{equation}\label{cccd}
\begin{gathered}
\xymatrix@C=6pc{\mathcal{A}_1 \ar[r]_{g_{12}}^{g_{21}} & \ar[l] \mathcal{A}_2 \ar[r]_{g_{23}}^{g_{32}} & \ar[l] \ldots ~\mathcal{A}_k \\
&\mathbf{C^\prime} \ar[ul]^{h_1} \ar[u]^{h_2} \ar[ur]_{h_k}& \\
\mathcal{A}_1 \ar[ur]^{f_1} \ar[r]_{g_{12}}^{g_{21}} & \ar[l] \mathcal{A}_2 \ar[u]_{f_2} \ar[r]_{g_{23}}^{g_{32}} & \ar[l] \ldots ~\mathcal{A}_k \ar[ul]_{f_k}
}
\end{gathered}
\end{equation}
\noindent
where the nodes $\mathcal{A}_i$ are Barwise-Seligman \cite{barwise:97} classifiers. The node $\mathbf{C^\prime}$ is a Barwise-Seligman classifier that is both the colimit of the incoming arrows $f_j$ and the limit of the outgoing arrows $h_j$, and all arrows are morphisms (``infomorphisms'') between such classifiers \cite{fg:19a}.  A Barwise-Seligman classifier implements a satisfaction relation $\Vdash_{\mathcal{A}}$ between ``tokens'' and ``types'' in some language.  Letting the tokens be bit strings in $\{ 0,1 \}^n$ and the types be bit strings in $\{ 0,1 \}^m$, we can consider $\Vdash_{\mathcal{A}}$ to be given by an $n \times m$ real matrix $P_{ij}$, where each element $p_{ij}$ represents the probability that the $i^{th}$ token belongs to the $j^{th}$ type \cite{fgm:21, fgm:22}; when all $p_{ij} \in \{ 0,1 \}$, binary classifiers as originally defined in \cite{barwise:97} are recovered.  Letting $\A$ and $\B$ be classifiers with tokens in $\rm{Tok}(\A)$ and $\rm{Tok}(\B)$ and types in $\rm{Typ}(\A)$ and $\rm{Typ}(\B)$, respectively, an {\em infomorphism} is a pair of maps $\overrightarrow{f}$ and $\overleftarrow{f}$ such that the following diagram commutes:

\begin{equation}\label{info-diagram-1}
\xymatrix@!C=3pc{\rm{Typ}(\A) \ar[r]^{\overrightarrow{f}}   & \rm{Typ}(\B) \ar@{-}[d]^{\Vdash_{\B}} \\
\rm{Tok}(\A) \ar@{-}[u]^{\Vdash_{\A}}  & \rm{Tok}(\B) \ar[l]_{\overleftarrow{f}}}
\end{equation}
Infomorphisms thus provide informational `semantic coherence' between classifiers, and are further amenable as such when the local logics of a (regular) theory are taken into account to create {\em logic infomorphisms} \cite[\S 12]{barwise:97} (reviewed in \cite{fg:19a}).
\footnote{From the category-theory perspective, Barwise-Seligman classifiers along with infomorphisms form a category isomorphic to that of Chu spaces with Chu morphisms, the latter category having been originally formulated by Barr and his student Chu \cite{barr:79}. Thereafter, significant developments in both the theory and applications of Chu spaces were undertaken by Pratt, mainly within theoretical computer science and conceptual modeling (see e.g. \cite{pratt:97,pratt:99a,pratt:00}, and the review in \cite{fg:19a}).}

Commutativity of CCCDs, i.e. diagrams of the form \eqref{cccd}, is guaranteed by the definition of $\mathbf{C^\prime}$ as both a limit and a colimit of infomorphisms to and from the $\A_i$.  Such diagrams can be arbitrarily elaborated by the addition of intermediate ``layers'' of classifiers with appropriate incoming and outgoing infomorphisms provided this commutativity condition is respected \cite{ffgl:22, fg:22}.  CCCDs are naturally interpreted as automorphisms $\{0, 1 \}^k \rightarrow \{0, 1 \}^k$ implemented by passage through a constraint network having the classifier $\mathbf{C^\prime}$ as its apex; they can be interpreted as implementing variational auto-encoders (VAEs) or arbitrary Bayesian networks as discussed in \cite{ffgl:22, fg:22}.  More generally, they can be taken to represent most types of functional (directed) graph networks along with their underlying quiver representations \cite{seigal:22} as applied in \cite{fgm:22}.

Non-commutativity of CCCDs, typically when $\mathbf{C^\prime}$ is undefinable for any hierarchical Bayesian network, for instance, is a compelling separate issue affording criteria for {\em intrinsic} or {\em quantum contextuality} as formulated by the results of \cite[\S 7]{fg:22} and \cite[\S 7.2]{fgm:22} (cf. \cite{abramsky:11,dzha:17b,gudder:19}). Essentially, such criteria involve {\em the non-existence of any ``globally'' definable (conditional) probability distribution across all possible observations.}  Regarding this separate issue, there is much to be said (and to be amplified elsewhere) in light of the non-commutativity results which are closely tied in with the development of the GHP and QRF formalism as presented here. For now, let us briefly comment upon the relevance: non-locality in QFT is a special case of quantum contextuality (see e.g. \cite{mermin:90,mermin:18,howard:14} for summaries of the Bell-Kochen-Specker theorems). In the quest for designing robust, fault-tolerant (FT), massive-scale performing quantum computers, quantum contextuality turns out to be an essential resource for quantum-speed up, encompassing such powerful techniques as magic state distillation (MSD) \cite{howard:14}, and related quantum computation by state injection (QCSI), as established for qubits relative to measurement-based quantum computation \cite{bermejo:17}.

\begin{figure}[H]
\centering
\includegraphics[width=15 cm]{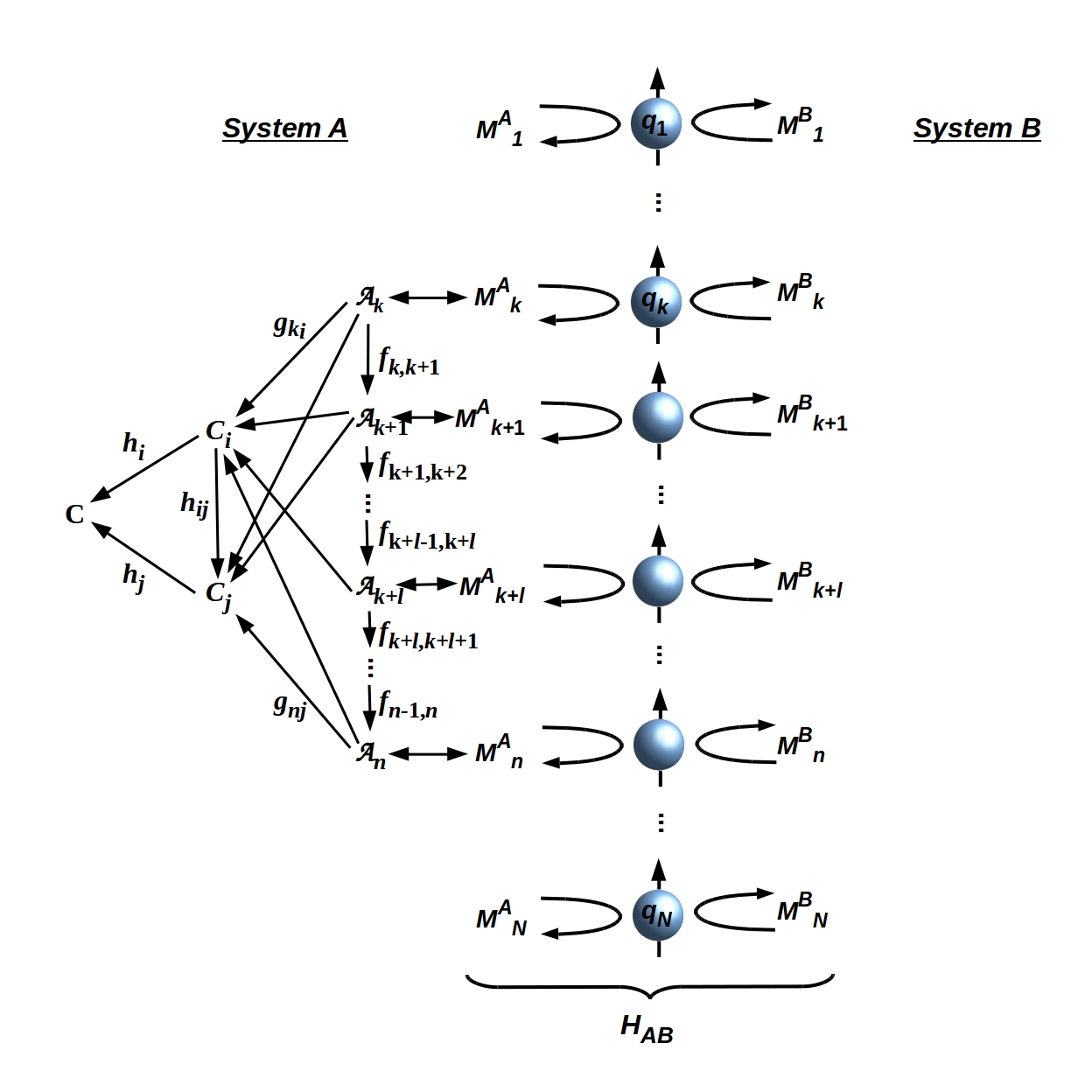}
\caption{Attaching a CCCD to a subset of measurement operators $M^A_k, \dots M^A_n$ by identifying the binary eigenvalues of the $M^A_i$ with binary inputs to the $\mathcal{A}_i$.  Only the incoming arrows are shown for simplicity; adding equivalent but reversed outgoing arrows completes the CCCD.  The CCCD specifies a function computed by the internal dynamics $H_A$, i.e. a QRF deployed by $A$.  Adapted from \cite{fgm:21} Fig. 3; CC-BY license.}
\label{CCCD-to-screen-fig}
\end{figure}

The ``raw data'' available to any QRF $Q$ implemented by an observer $A$ are the eigenvalues $+1$ or $-1$ returned by some subset of the operators $M^A_i$.  We can, therefore, represent any such $Q$ as a CCCD ``attached'' to the boundary $\mathscr{B}$ as shown in Fig. \ref{CCCD-to-screen-fig}.  The measured bits are prepared by the action of the ``world'' $B$'s corresponding operators $M^B_i$; the CCCD acts back on the measured bits to prepare them for subsequent measurement by $B$, preserving the symmetric, cyclic interaction required by Eq. \eqref{ham}.

The classifier $\A_i$ accepting an input of $+1$ or $-1$ from the measurement operator $M^A_i$ can be defined to execute any function $\varphi: \{1,-1 \} \rightarrow [0,1]$, i.e. to assign any probability value.  Operationally, therefore, the classifier $\A_i$ acts as the local $z$ axis with respect to which the qubit $q_i$ on which $M^A_i$ acts is measured or prepared.  This is in fact obvious: the local $z$ axis must itself be a physically-implemented one-bit QRF.  Choosing the local $z$ axis is, as discussed earlier, equivalent to choosing the basis in which the $M^A_i$ are expressed.  Free choice of basis for the $M^A_i$ implies, therefore, free choice of QRFs; a QRF acting on the outputs of a subset of operators $M^A_i \dots M^A_k$ effectively sets the local basis for these operators.  Free choice of QRFs enables observers to treat the data encoded on different components of their boundaries differently, and hence to distinguish ``systems of interest'' from their surrounding environments.  Formally, implementing a QRF $Q$ breaks the $S_N$ swap symmetry of $\mathscr{B}$ by assigning the qubits measured by the $M^A_i \dots M^A_k$ the functional role of ``inputs'' to $Q$.  Free choice moreover entails, consistent with the discussion in \S \ref{choice} above, that $A$ and $B$ can choose different QRFs, and hence both process and prepare states of $\mathscr{B}$ in different ways.  Sharing QRFs across $\mathscr{B}$ induces entanglement \cite{fgm:21, ffgl:22}, a point to which we will return in \S \ref{entanglement} below.

\subsection{Observable systems and their pointer states are defined by QRFs} \label{systems}

\begin{figure}[H]
\centering
\includegraphics[width=13 cm]{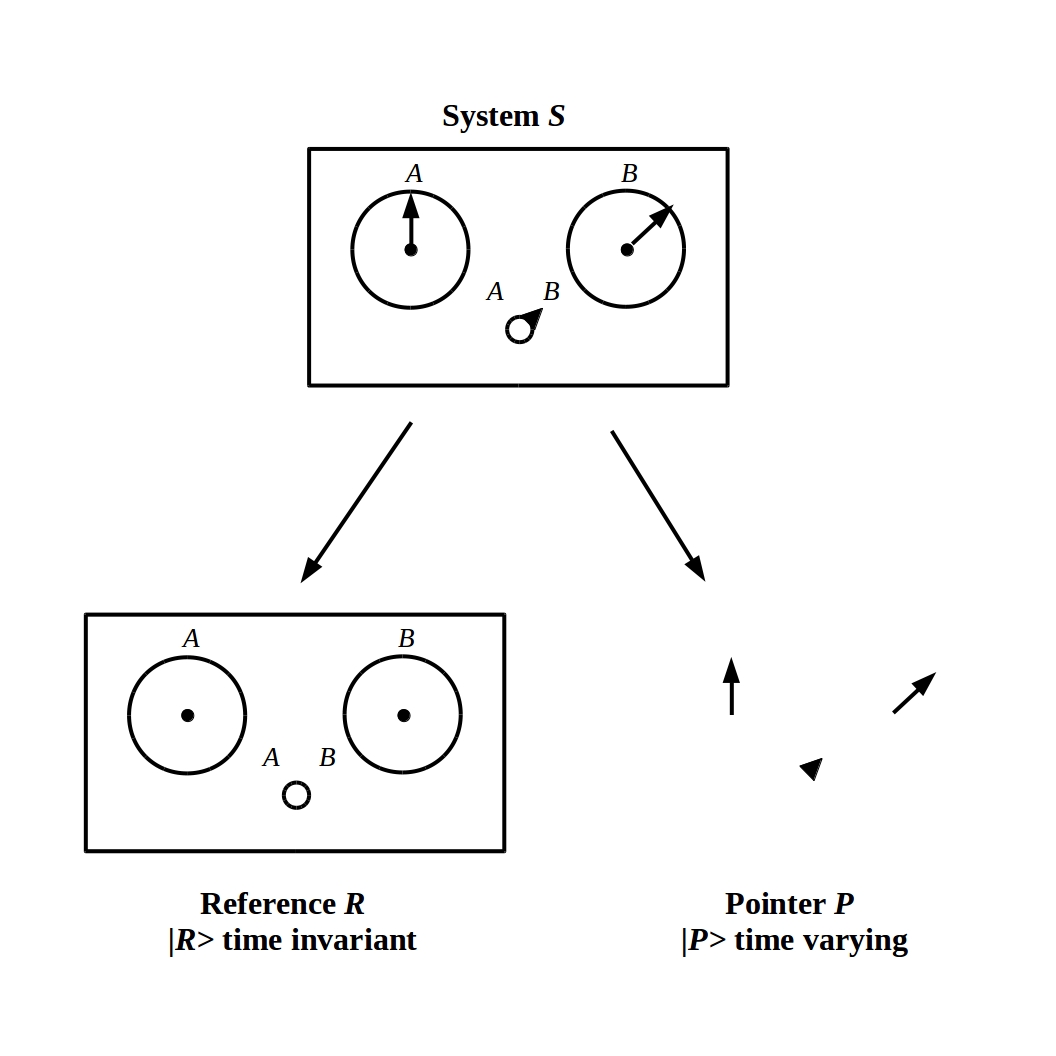}
\caption{Identifying a system $S$ requires identifying some proper component $R$ that maintains a constant state $\vert R \rangle$ (or density of time-averaged samples $\rho_R$) as the ``pointer'' state $\vert P \rangle$ (or density of time-averaged samples $\rho_P$) of interest varies.  Adapted from \cite{fg:20} Fig. 2, CC-BY license.}
\label{ref-vs-pointer-fig}
\end{figure}

Consider a simple and canonical case of measurement: at some time $t_i$, a human observer $A$ measures the time-dependent pointer state $|P \rangle$, or at lower resolution a pointer-state density $\rho_P$, of some system $S$ of interest, e.g. an item of laboratory apparatus.  For $S$ to be measureable, it must be part of $A$'s observable ``environment'' $E$.  Prior to measuring $|P \rangle$, the observer must {\em identify} $S$, distinguishing it from the rest of $E$, including other items of apparatus that are not $S$ as well as the rest of the laboratory and beyond.  This identification process cannot depend on the state $|P \rangle$, which is of interest as measurement target precisely because it is both unknown and time-dependent.  Identification instead depends on some ``reference'' degrees of freedom of $S$, e.g. its size, shape, color, labeling, or location, that are time invariant, i.e. that maintain some constant state $|R \rangle$ or state density $\rho_R$.  Figure \ref{ref-vs-pointer-fig} illustrates this commonplace scenario.

The data specifying the state $|E \rangle$ of the undifferentiated environment, and the state $|S \rangle = |R \rangle |P \rangle$ of the identified system of interest to $A$, can only be proper subsets or sectors of the data encoded on $\mathscr{B}$ and measured by the $M^A_i$.  These sectors must, moreover, be disjoint from the thermodynamic sector $F$ from which $A$ extracts sufficient free energy to support any classical information processing.  We can, therefore, identify proper subsets of operators $M^E_i$, $M^R_i$, and $M^P_i$, dropping the superscript $A$ to simplify the notation.  The outputs of these subsets of operators are processed by QRFs that can be labeled $E$, $R$, and $P$ without ambiguity; the sectors specifying the states $|E \rangle$, $|R \rangle$, and $|P \rangle$ are simply the domains of the operators $M^E_i$, $M^R_i$, and $M^P_i$, respectively, and hence of these QRFs.  The fact that $S$ is part of $E$ requires that $\{ M^R_i \}, \{ M^P_i \} \subset \{ M^E_i \}$ and, therefore, that $R$ and $P$ are proper components of $E$.  Co-measurability of $R$ and $P$ requires that $R$ and $P$ are decoherent both from each other and from the remainder of $E$ \cite{fm:20, fg:20, fgm:21}.

\begin{figure}[H]
\centering
\includegraphics[width=13 cm]{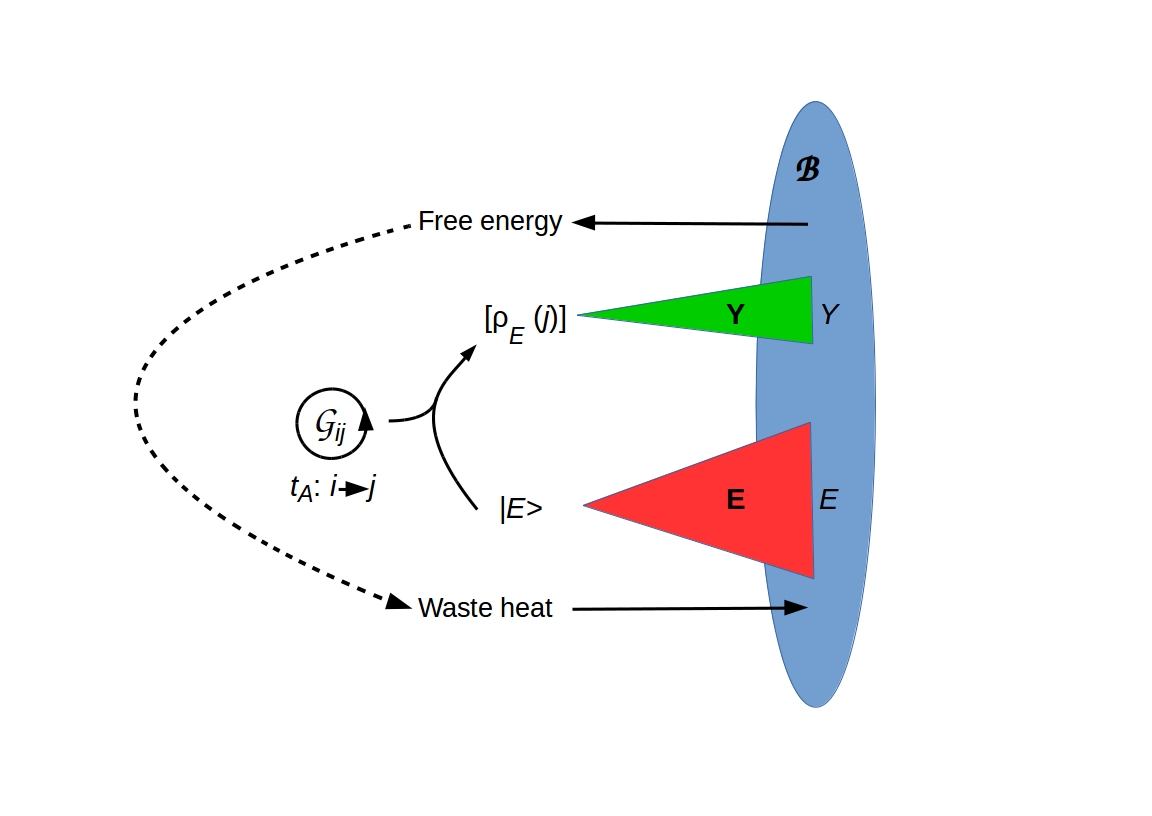}
\caption{Cartoon illustration of QRFs required to observe and write a readable memory of an environmental state $\vert E \rangle$.  The QRFs $\mathbf{E}$ and $\mathbf{Y}$ read the state from $E$ and write it to the memory $Y$ respectively.  Any identified system $S$ must be part of $E$.  The clock $\mathcal{G}_{ij}$ is a time QRF that defines the time coordinate $t_A$.  The dashed arrow indicates the observer's thermodynamic process that converts free energy obtained from the unobserved sector $F$ of $\mathscr{B}$ to waste heat exhausted through $F$.}
\label{mem-write-fig}
\end{figure}

Developing a model of the behavior of $P$ that enables predictions of future states requires, at minimum, recording measurements of $|P \rangle$ taken at multiple times to some classical memory.  Typically at least some of the ambient background conditions encoded in $| E \rangle$ as a whole are also recorded.  Like any physical system with which an observer interacts, a memory $Y$ must be identified before being written on or read from.  The considerations adduced above for any system $S$ thus apply equally to any memory $Y$.  The basic elements of a quantum theory of measurement can, therefore, be depicted as in Fig. \ref{mem-write-fig}.  They include not only the QRFs discussed here and the thermodynamic flows that power them, but also a time QRF that provides a measure of duration between writes to memory and hence an effective time stamp.  This timekeeping system will be discussed further in \S \ref{time} below.

The ``picture'' of measurement illustrated in Fig. \ref{mem-write-fig} differs in significant ways from that introduced by von Neumann \cite{vonN:55} and reproduced in most textbooks.  Most obviously, it treats the observer as a physical system with a particular functional architecture, not as an abstraction.  It enforces, via the GHP, the requirement of separability between the observer and the world being observed; without this, the observer lacks a conditionally-independent state and the idea of ``measurement'' becomes meaningless.  The GHP also restricts the observer's access to the ``bulk'' degrees of freedom of the world: the operators $H^A_i$ act not on the Hilbert space $\mathcal{H}_B$ of $B$, but on the much smaller effective Hilbert space $H^A_{\mathscr{B}}$ of the boundary $\mathscr{B}$.  This renders all observed ``systems'' observer-relative and hence ``personal'' in the sense emphasized by QBists \cite{fuchs:10, mermin:18}.  Observed ``systems'' here include memory devices and, significantly, other observers; hence the theory of measurement requires a {\em physical} theory of classical communication that has yet to be fully developed \cite{fgm:21, ffgl:22}.  Perhaps most subtly, Fig.~\ref{mem-write-fig} involves no assumption of a background spatial embedding.  It treats 3d space as a QRF that $A$ may or may not be capable of deploying.  Hence Fig.~\ref{mem-write-fig} is consistent with approaches to QG in which spacetime is fully emergent from underlying informational or other processes.  The field theory that naturally follows from Fig.~\ref{mem-write-fig} is, therefore, a topological quantum field theory (TQFT), not a QFT on a background spacetime.

\subsection{Sequential measurements induce TQFTs} \label{TQFTs}

We have shown in \cite{fgm:22} that given the GHP, sequential measurements of any sector $S$ of a holographic screen $\mathscr{B}$ induce a TQFT on $S$.  We also show how this TQFT can be realized as a quantum topological neural network (TQNN), a generalized representation of a standard deep-learning system \cite{marciano:22}.  Here we briefly summarize the main result and mention some of its consequences, referring readers to \cite{fgm:22} for details.

A TQFT can be represented as a functor from the category of cobordisms to the category of Hilbert spaces \cite{atiyah:88, quinn:95}.  We prove in \cite{fgm:22} that any QRF can be represented as a CCCD, and then construct a category with CCCDs as objects and morphisms of CCCDs, which must by definition respect the commutativity of CCCDs as diagrams, as morphisms.  The category is, effectively, a category of QRFs, in which the morphisms represent sequential choices of QRF to be applied to the data encoded on some sector $S$.  We show that all such choices can be represented by one of two diagrams.  Using the compact notation

\begin{equation} \label{QRF-1}
\begin{gathered}
\begin{tikzpicture}
\draw [thick] (0,0) -- (2,1) -- (2,-1) -- (0,0);
\node at (1.3,0) {S};
\end{tikzpicture}
\end{gathered}
\end{equation}
\noindent
to represent a QRF $S$, we can represent measurements of a physical situation in which one system divides into two, possibly entangled, systems with a diagram of the form

\begin{equation} \label{QRF-2}
\begin{gathered}
\begin{tikzpicture}
\node at (0,0) {$S$};
\draw [thick] (0.2,0) -- (1,1) -- (2,1.5) -- (2,0.5) -- (1,1);
\node at (1.7,1) {$S_1$};
\draw [thick] (0.2,0) -- (1,-1) -- (2,-0.5) -- (2,-1.5) -- (1,-1);
\node at (1.7,-1) {$S_2$};
\end{tikzpicture}
\end{gathered}
\end{equation}
\noindent
Parametric down-conversion of a photon exemplifies this kind of process.  The reverse process can be added to yield:

\begin{equation} \label{flow-1}
\begin{gathered}
\begin{tikzpicture}
\node at (0,0) {$S$};
\draw [thick] (0.2,0) -- (1,1) -- (2,1.5) -- (2,0.5) -- (1,1);
\node at (1.7,1) {$S_1$};
\draw [thick] (0.2,0) -- (1,-1) -- (2,-0.5) -- (2,-1.5) -- (1,-1);
\node at (1.7,-1) {$S_2$};
\draw [thick] (-2.7,0) -- (-1.7,0.5) -- (-1.7,-0.5) -- (-2.7,0);
\node at (-2,0) {$S$};
\draw [thick, ->] (-1.5,0) -- (-0.3,0);
\draw [thick, ->] (2.2,0) -- (3.4,0);
\draw [thick] (3.6,0) -- (4.6,0.5) -- (4.6,-0.5) -- (3.6,0);
\node at (4.3,0) {$S$};
\end{tikzpicture}
\end{gathered}
\end{equation}
\noindent
Diagram \eqref{flow-1} represents a relabelling of subsets of the base-level classifiers that act on the sector $S$:

\begin{equation} \label{class-flow-1}
\underbrace{\mathcal{A}_1, \mathcal{A}_2, \dots \mathcal{A}_m}_{S} \rightarrow \underbrace{\mathcal{A}_1, \dots \mathcal{A}_i,}_{S_1} \underbrace{\mathcal{A}_{i+1}, \dots \mathcal{A}_m}_{S_2} \rightarrow \underbrace{\mathcal{A}_1, \mathcal{A}_2, \dots \mathcal{A}_m}_{S}
\end{equation}
\noindent
In the second type of sequential measurement process, the pointer-state QRF $P$ is replaced with an alternative QRF $Q$ with which it does not commute.  Sequences in which position and momentum, $s_z$ and $s_x$ are measured alternately are examples.  These can be represented by the diagram

\begin{equation} \label{flow-2}
\begin{gathered}
\begin{tikzpicture}
\node at (0,0) {$\mathbf{S}$};
\draw [thick] (0.2,0) -- (1,1) -- (2,1.5) -- (2,0.5) -- (1,1);
\node at (1.7,1) {$\mathbf{P}$};
\draw [thick] (0.2,0) -- (1,-1) -- (2,-0.5) -- (2,-1.5) -- (1,-1);
\node at (1.7,-1) {$\mathbf{R}$};
\draw [thick] (-2.7,0) -- (-1.7,0.5) -- (-1.7,-0.5) -- (-2.7,0);
\node at (-2,0) {$\mathbf{S}$};
\draw [thick, ->] (-1.5,0) -- (-0.3,0);
\draw [thick, ->] (2.2,0) -- (3.4,0);
\node at (3.7,0) {$\mathbf{S}$};
\draw [thick] (3.9,0) -- (4.7,1) -- (5.7,1.5) -- (5.7,0.5) -- (4.7,1);
\node at (5.4,1) {$\mathbf{Q}$};
\draw [thick] (3.9,0) -- (4.7,-1) -- (5.7,-0.5) -- (5.7,-1.5) -- (4.7,-1);
\node at (5.4,-1) {$\mathbf{R}$};
\draw [thick, ->] (5.9,0) -- (7.1,0);
\draw [thick] (7.3,0) -- (8.3,0.5) -- (8.3,-0.5) -- (7.3,0);
\node at (8,0) {$\mathbf{S}$};
\end{tikzpicture}
\end{gathered}
\end{equation}
\noindent
Again this can be written as a relabeling of classifiers, leaving the pointer-state classifiers that are traced over when measuring only the reference component $R$ for system identification implicit, as:

\begin{equation} \label{class-flow-2}
\underbrace{\mathcal{A}_1, \mathcal{A}_2, \dots \mathcal{A}_k}_{R} \rightarrow \underbrace{\mathcal{A}_1, \dots \mathcal{A}_k,}_{R} \underbrace{\mathcal{A}_{k+1}, \dots \mathcal{A}_m}_{P} \rightarrow \underbrace{\mathcal{A}_1, \dots \mathcal{A}_k,}_{R} \underbrace{\tilde{\mathcal{A}}_{k+1}, \dots \tilde{\mathcal{A}}_m}_{Q} \rightarrow \underbrace{\mathcal{A}_1, \mathcal{A}_2, \dots \mathcal{A}_k}_{R}
\end{equation}
\noindent
where the notation $\tilde{\mathcal{A}}_{l}$ indicates that $\mathcal{A}_{l}$ has been rewritten in a rotated measurement basis, e.g. $s_z \rightarrow s_x$ or $x \rightarrow p= m\, (\partial x/ \partial t)$.  As both $P$ and $Q$ must commute with $R$, the commutativity requirements for $S$ are satisfied.

Measurement sequences of the form of Diagram \eqref{flow-1} can be mapped to cobordisms of the form

\begin{equation} \label{CCCD-to-Cob}
\begin{gathered}
\begin{tikzpicture}[every tqft/.append style={transform shape}]
\draw[rotate=90] (0,0) ellipse (2.5cm and 1 cm);
\node[above] at (0,1.7) {$\mathscr{B}$};
\node at (-0.4,0) {$S$};
\begin{scope}[tqft/every boundary component/.style={draw,fill=green,fill opacity=1}]
\begin{scope}[tqft/cobordism/.style={draw}]
\begin{scope}[rotate=90]
\pic[tqft/cylinder, name=a];
\pic[tqft/pair of pants, anchor=incoming boundary 1, name=b, at=(a-outgoing boundary 1)];
\end{scope}
\end{scope}
\end{scope}
\draw[rotate=90] (0,-4) ellipse (2.5cm and 1 cm);
\node[above] at (4,1.7) {$\mathscr{B}$};
\node at (2,0.7) {$\mathscr{S}$};
\node at (4.4,1.1) {$S_1$};
\node at (4.4,-1.1) {$S_2$};
\draw [thick, <-] (0,-2.7) -- (0,-3.8);
\draw [thick, <-] (4,-2.7) -- (4,-3.8);
\draw [thick] (-0.5,-5.3) -- (0.5,-4.8) -- (0.5,-5.8) -- (-0.5,-5.3);
\node at (0.2,-5.3) {$\mathbf{S}$};
\draw [thick] (3.5,-4.5) -- (4.5,-4) -- (4.5,-5) -- (3.5,-4.5);
\node at (4.2,-4.5) {$\mathbf{S_1}$};
\draw [thick] (3.5,-6.1) -- (4.5,-5.6) -- (4.5,-6.6) -- (3.5,-6.1);
\node at (4.2,-6.1) {$\mathbf{S_2}$};
\draw [thick] (3.5,-4.5) -- (2.5,-5.3) -- (3.5,-6.1);
\draw [thick, ->] (0.7,-5.3) -- (2.3,-5.3);
\node at (-0.5,-3.3) {$\mathfrak{F}(i)$};
\node at (4.5,-3.3) {$\mathfrak{F}(k)$};
\node at (1.5,-5.6) {$\mathscr{F}$};
\end{tikzpicture}
\end{gathered}
\end{equation}

while sequences of the form of Diagram \eqref{flow-2} can be mapped to cobordisms of the form:

\begin{equation} \label{CCCD-to-Cob-2}
\begin{gathered}
\begin{tikzpicture}[every tqft/.append style={transform shape}]
\draw[rotate=90] (0,0) ellipse (2.5cm and 1 cm);
\node[above] at (0,1.7) {$\mathscr{B}$};
\node at (-0.5,1.1) {$P$};
\node at (-0.5,-1.1) {$R$};
\begin{scope}[tqft/every boundary component/.style={draw,fill=green,fill opacity=1}]
\begin{scope}[tqft/cobordism/.style={draw}]
\begin{scope}[rotate=90]
\pic[tqft/reverse pair of pants, at={(-1,0)}, name=a];
\pic[tqft/pair of pants, anchor=incoming boundary 1, name=b, at=(a-outgoing boundary 1)];
\end{scope}
\end{scope}
\end{scope}
\draw[rotate=90] (0,-4) ellipse (2.5cm and 1 cm);
\node[above] at (4,1.7) {$\mathscr{B}$};
\node at (2,0.7) {$\mathscr{S}$};
\node at (4.4,1.1) {$Q$};
\node at (4.4,-1.1) {$R$};
\draw [thick, <-] (0,-2.7) -- (0,-3.8);
\draw [thick, <-] (4,-2.7) -- (4,-3.8);
\draw [thick] (-0.5,-4.5) -- (0.5,-4) -- (0.5,-5) -- (-0.5,-4.5);
\node at (0.2,-4.5) {$\mathbf{P}$};
\draw [thick] (-0.5,-6.1) -- (0.5,-5.6) -- (0.5,-6.6) -- (-0.5,-6.1);
\node at (0.2,-6.1) {$\mathbf{R}$};
\draw [thick] (-0.5,-4.5) -- (-1.6,-5.3) -- (-0.5,-6.1);
\node at (-1.8,-5.3) {$\mathbf{S}$};
\draw [thick] (3.5,-4.5) -- (4.5,-4) -- (4.5,-5) -- (3.5,-4.5);
\node at (4.2,-4.5) {$\mathbf{Q}$};
\draw [thick] (3.5,-6.1) -- (4.5,-5.6) -- (4.5,-6.6) -- (3.5,-6.1);
\node at (4.2,-6.1) {$\mathbf{R}$};
\draw [thick] (3.5,-4.5) -- (2.5,-5.3) -- (3.5,-6.1);
\draw [thick, ->] (0.7,-5.3) -- (2.3,-5.3);
\node at (-0.5,-3.3) {$\mathfrak{F}(i)$};
\node at (4.5,-3.3) {$\mathfrak{F}(k)$};
\node at (1.5,-5.6) {$\mathscr{F}$};
\end{tikzpicture}
\end{gathered}
\end{equation}

In either case, $\mathfrak{F}: \mathbf{CCCD} \rightarrow \mathbf{Cob}$ is the required functor from the category $\mathbf{CCCD}$ of CCCDs to the category of $\mathbf{Cob}$ finite cobordisms.  In general, we can state:

\begin{theorem}[\cite{fgm:22} Thm. 1] \label{thm2}
For any morphism $\mathscr{F}$ of CCCDs in $\mathbf{CCCD}$, there is a cobordism $\mathscr{S}$ such that a diagram of the form of Diagram \eqref{CCCD-to-Cob} or \eqref{CCCD-to-Cob-2} commutes.
\end{theorem}
\noindent
referring to \cite{fgm:22} for the proof.

Theorem \ref{thm2} has a number of immediate consequences, chief among which is that any effective field theory (EFT) defined on $S$ must be gauge invariant \cite{addazi:21, fgm:22}.  The GHP, therefore, not only generates a default physical theory -- a TQFT -- of any observable system, but strongly restricts any geometrization of that theory.  Indeed the results obtained in \cite{fgm:22} strongly suggest that observable spatial geometry, including the Minkowski metric and the Einstein equations of GR, is induced by symmetries of the QRFs employed to identify observable systems as such over time.  If this proves to be the case, it will reconceptualize ``space'' as a quantum informational structure even at macroscopic scales.

\section{The GHP provides local definitions of entropy and time} \label{time}

\subsection{Operations on $\mathscr{B}$ implement Wick rotations}

The GHP generalizes the Bekenstein area law \cite{bekenstein:73} to a statement applicable to any boundary $\mathscr{B}$ implementing an interaction $H_{AB}$ between finite, separable systems, i.e. an interaction that can be written as Eq. \eqref{ham}:

\begin{quote}
A boundary $\mathscr{B}$ implementing an interaction $H_{AB}$ between finite, separable systems has thermodynamic entropy $S(\mathscr{B}) = N$, where $N$ is the dimension of $H_{AB}$.
\end{quote}
\noindent
Geometrization of $\mathscr{B}$ then requires $N \leq A(\mathscr{B})/4$ as discussed in \S \ref{exchange} above.  The $S(\mathscr{B})$ is thus conceptually an entropy as defined by Shannon \cite{shannon:48}: the width of a classical communication channel.

Unlike $S(\mathscr{B})$, the thermodynamic entropy $S(B)$ of the physical system $B$ is neither specified nor rerstricted by $H_{AB}$.  From $A$'s perspective, however, $B$ is a source of both usable free energy and classical information as illustrated in Fig.~\ref{mem-write-fig}.  Relative to $A$, therefore, $S(B)$ cannot decrease, i.e. $B$ cannot become a free-energy or classical-information sink.  The 2nd Law thus holds for $A$, independently of either $H_B$ or of any details of $A$'s predictive models, if any, of $B$'s behavior.  The informational symmetry of $\mathscr{B}$ guarantees the same is true for $B$: relative to $B$, $S(A)$ cannot decrease.  This observer-relativity of thermodynamic entropy has previously been emphasized by Tegmark \cite{tegmark:12}.

As discussed in \S \ref{systems} above, the idea of sequential measurement, and hence the idea of recordable time, is only physically meaningful for observers able to write data irreversibly to a classical memory.  The action of writing to a memory sector $Y$ defines an $A$-specific, local time QRF $t_A$ as illustrated in Fig.~\ref{mem-write-fig}.  The most natural unit of $t_A$ is the minimal time to write one bit, to which time-energy complementary gives a minimum value of $h / [\mathrm{ln} 2 (K_B T^A)]$, with $h$ being the Planck's constant.  The bit-counting process can be implemented by an operator $\mathcal{G}_{ij}$ that advances $t_A$ by one unit $i \rightarrow j$; formally, $\mathcal{G}_{ij}$ is a groupoid element \cite{fg:20, fgm:21}.  The rate at which $A$'s ``clock'' $\mathcal{G}_{ij}$ ``ticks'' is determined by $A$'s thermodynamic efficiency.

\begin{figure}[H]
\centering
\includegraphics[width=10 cm]{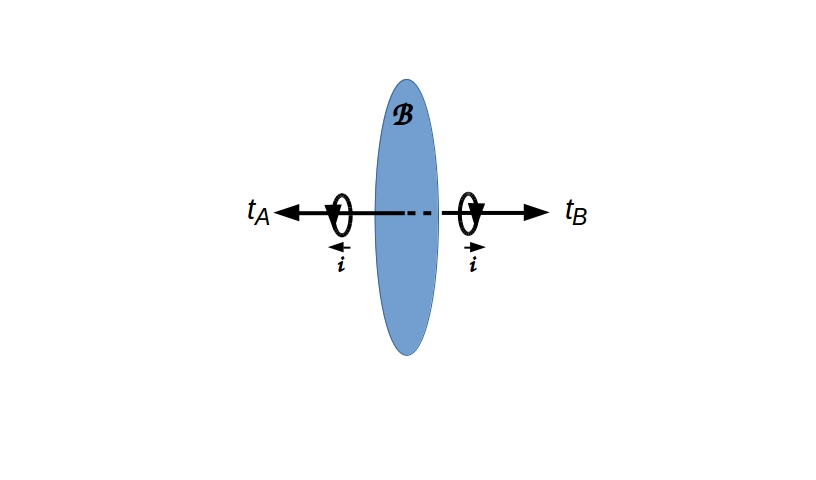}
\caption{Local times $t_A$ and $t_B$ flow in opposite directions across $\mathscr{B}$.    Each ``write'' or ``prepare'' operation on $\mathscr{B}$ thus implements a Wick rotation $\protect\overrightarrow{\imath}$ of the local time, with a total energetic cost of a combined write-read of at least $\hbar / \mathrm{ln} 2 (K_B T^A)$.}
\label{time-wick-fig}
\end{figure}

The local time QRF $t_A$ is clearly entropic: it counts recordable information received from $B$ and hence increments as ($A$-relative) $S(B)$ increments (see \cite{rovelli:21} for a similar account of entropic time). Thus it is natural to interpret the measured time $t_A$ as ``flowing'' with the passage of information from $B$ to $A$.  The informational symmetry of $\mathscr{B}$ allows us to represent $t_B$ in the same way, as illustrated in Fig.~\ref{time-wick-fig}.  We can, therefore, see the GHP as giving a physical meaning to the Wick rotation \cite{baez:20}, namely to the prescription that ``inverse temperature is imaginary time'': a measurement operation performed by $A$ on a qubit received from $B$ induces the ``collapse'' of the qubit into a certain eigenstate $\varepsilon $, namely $|q \rangle_t \rightarrow e^{-\imath \varepsilon t} | \varepsilon \rangle \sim | \varepsilon \rangle $, where we can write  $| \varepsilon \rangle $ as a pure state because its time-phase dependence is not observable. The pure state hence recovered by the QRF of $A$ represents the element of classical information that is processed thermodynamically on $\mathscr{B}$, hence subjected to a thermodynamic distribution $e^{-\varepsilon/(K_B T)}$. Encoding of information can be seen therefore as a Wick rotation $\imath  \varepsilon  t \rightarrow \varepsilon/(K_B T)$. A further such process of ``reading'' performed by $B$ can be understood as a backward evolution in time of the qubit before irreversible encoding happens, or as an evolution of the missing (virtual, because of irreversible encoding has happened) qubit of energy $-\varepsilon$. This virtual evolution of a ``hole''-like qubit would in turn correspond to a second Wick rotation, with same axis of rotation, reverting the axis of time. In other words, each operation on a qubit of $\mathscr{B}$ rotates the local time vector by $\overrightarrow{\imath}$, so a combined cross-$\mathscr{B}$ write-read operation in either the $B$-to-$A$ or $A$-to-$B$ direction implements $\overrightarrow{\imath}$ twice and reverses the local time direction. Energy-momenta and angular momenta are not conserved during the encoding process: the energetic cost of the rotation is rather understood in terms of an irreversible bit encoding, i.e. at least $ \mathrm{ln} 2 (K_B T^A)$.

Figure \ref{time-wick-fig} shows that any system $A$ satisfying the separability conditions required by the GHP can be viewed as interacting with its own future.  Hence any decomposition $U = AB$ that respects separability can be viewed as decomposing $\mathcal{H}_U$ along a temporal boundary.  When the boundary $\mathscr{B}$ is given the geometry shown in Fig. \ref{qubit-fig} and hence required to respect covariance, this temporal decomposition has an obvious interpretation: any separable system interacts exclusively with its own future light cone.  The implementation of Wick rotation by actions on $\mathscr{B}$ is thus intimately tied to the role of gauge bosons as information carriers and hence to the Minkowski metric as a representation of the time dependence of information flow.

\subsection{QRF sharing induces entanglement} \label{entanglement}

Figures \ref{mem-write-fig} and \ref{time-wick-fig} enable a simple and intuitive understanding of the relation between free choice and separability, and of the approach to entanglement as these are violated.  Suppose $A$ and $B$ implement QRFs $E^A$, $E^B$ and $Y^A$, $Y^B$, respectively, such that $E^A$ and $E^B$ compute the same function $\varphi$ and $Y^A$ and $Y^B$ compute the same function $\psi$.  As arbitrarily many distinct physical systems can compute any given function, this is merely an assumption of shared classical information processing.  Now assume that $dom(E^A) = ~dom(E^B)$ and that $dom(Y^A) = ~dom(Y^B)$, with $dom$ denoting a function's domain, i.e. that each pair of operators acts on a shared subset of encoded bits.  This is a quantum assumption, as it is an assumption about how $E^A$, $E^B$ and $Y^A$, $Y^B$ are implemented by the internal Hamiltonians $H_A$ and $H_B$, respectively.  It does not, however, determine the time dependence of the states $|A \rangle$ or $|B \rangle$; in particular, it does not force $A$'s data writes to $dom(Y^A) = ~dom(Y^B)$ to synchronize with $B$'s data writes.  Adding the assumption that $t_A$ and $t_B$ have equal periods, however, does force such synchrony.  With this synchronization assumption, $A$ and $B$ update each other's memory sectors on each cycle.  Components of $|A \rangle$ and $|B \rangle$ that are memory dependent are, in this case, no longer conditionally independent.  Hence the joint state $|AB \rangle$ is no longer separable.

Let $\rho_A = \tr_B \vert \psi \rangle \langle \psi \vert$ be the reduced density matrix of $A$ obtained by taking a partial trace over $B$ of the total density matrix $\rho = \vert \psi \rangle \langle \psi \vert$ of the joint system $AB$.  Recall that the (von Neumann) entanglement entropy $\mathcal{S}(A)$ of $A$ in the bipartite decomposition $AB$ is given by (see e.g. \cite{nielsen:00,ryu:06}):
\begin{equation}\label{entangle-1}
\mathcal{S}(A)  = - \tr \rho_A \log \rho_A
\end{equation}
If the joint state $|AB \rangle$ is no longer separable, the entanglement entropy $\mathcal{S}(AB)$ is nonzero.  In the situation described above, we can localize this entanglement entropy to the particular sector $dom(Y^A) = ~dom(Y^B)$ of the decompositional boundary $\mathscr{B}$; in this case its maximum value is the dimension of this sector, i.e. $\mathcal{S}(AB)_{max} \sim dim(dom(Y^A)) = dim(dom(Y^B))$.
\footnote{In \cite[\S1]{headrick:14} entanglement entropy in the context of the AdS/CFT correspondence can be seen as a {\em wedge observable}, meaning that if spatial regions $A$
and $A'$ share the domain of dependence, then $dom(A) = ~dom(A')$, and they have the same entanglement entropy $\mathcal{S}_{AB} = \mathcal{S}_{A'B'}$ since the corresponding density matrices $\rho_A$ and $\rho_{A'}$ are unitarily related \cite{casini:04}. }

Requiring that $Y^A$ and $Y^B$ compute the same function $\psi$, that $dom(Y^A) = ~dom(Y^B)$, and that $t_A$ and $t_B$ have equal periods is, effectively, requiring that $Y^A$ and $Y^B$ are the same QRF.  We have previously shown \cite{fgm:21} that QRF replication across a boundary $\mathscr{B}$ is forbidden by the no-cloning theorem \cite{wooters:82}.  Briefly, the information on $\mathscr{B}$ does not determine $|A \rangle$, so it is insufficient for $B$ to replicate $|A \rangle$, and hence insufficient for $B$ to replicate any QRF state $|Q \rangle$ that is a component of $|A \rangle$ \cite{fm:19, fgm:21}.  No cloning of unknown states is, in this and indeed in any case, a straightforward consequence of the nonfungibility of quantum information.  The above discussion exemplifies this: the assumptions that $dom(Y^A) = ~dom(Y^B)$ and that $t_A$ and $t_B$ have equal periods can be made only as {\em a priori} stipulations, as neither of these conditions can be inferred from the data encoded on $\mathscr{B}$.  The GHP, therefore, provides a mechanism that enforces no-cloning by restricting the transfer of information between $A$ and $B$ to the information encoded on $\mathscr{B}$.

These results, together with those of the previous sections, provide a novel characterization of some standard concepts, including AdS/CFT as developed by \cite{maldecena:98} and others, while working throughout with both the bulk distribution and the boundary degrees of freedom specified in terms of binary qubits.  There are a number of related results. For instance, Ryu and Takayanagi \cite{ryu:06} start from the AdS/CFT correspondence to develop an HP-motivated derivation of entanglement entropy in $d+1$ dimensional CFT as obtainable from the area of a $d$-dimensional minimal surface $\Sigma$ in $\mathrm{AdS}_{d+2}$, a result analogous to the Bekenstein-Hawking formula for BH entropy. Related is the proposal of Lee \cite{lee:11} that entanglement arises from the HP, in so far that all the information specifying the physics bulk bits can be described in terms of qubits on the holographic boundary; a result which appears consonant with our development of ideas as presented above. An interesting direction of research hinges on substituting the AdS bulk with either a TQFT or an extended version of it, for instance an extended $BF$ theory. An attempt for an exact holographic mapping, with emergent space-time geometry recovered along the lines of \cite{Qi:2013caa}, has been investigated in \cite{Han:2016xmb}, where a relation between loop quantum gravity and tensor networks has been explored accounting for bulk-boundary duality and holographic entanglement entropy. From a genuine TQFT perspective, it would be tempting to analyze the connection between AdS bulk and $BF$ theories, with holographic boundaries provided by Chern-Simons theories with punctures. In lower dimensions, interesting results have been reported in $2d$ dimensional spaces with holographic boundaries involving the SYK model \cite{Mertens:2017mtv,Blommaert:2018oro}. More generally, Wootters \cite{wootters:01} reviews entanglement of pure states (informationally reversible), mixed states (irreversible, since in creating a mixed state from a pure state some information is forsaken), and entanglement of formation which is intimately tied to the notion of computational {\em concurrency} where explicit formulas are available.

\section{Areal elements are scattering centers} \label{scattering}

\begin{figure}[H]
\centering
\includegraphics[width=12 cm]{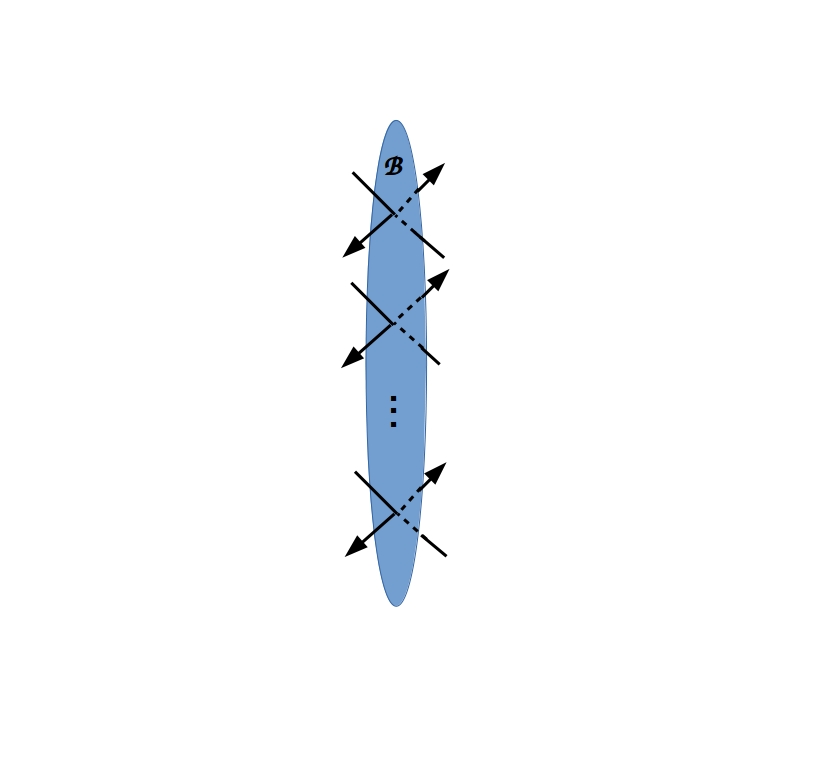}
\caption{Fig. \ref{qubit-screen-fig} re-drawn to represent areal elements as scattering centers.}
\label{scattering-fig}
\end{figure}

Replacing curved arrows with angled arrows and an explicit qubit array with $\mathscr{B}$, we can re-draw Fig.~\ref{qubit-screen-fig} as Fig.~\ref{scattering-fig}.  Viewing the arrows as depicting information flow as before, the areal elements of $\mathscr{B}$ function as scattering centers.  Inspecting this scattering diagram, two things are immediately apparent:

\begin{itemize}
\item From either $A$'s or $B$'s perspective, transmitting information across $\mathscr{B}$ is indistinguishable from scattering information off $\mathscr{B}$.
\item Comparing Fig.~\ref{time-wick-fig} and Fig.~\ref{scattering-fig}, scattering information off $\mathscr{B}$ reverses its temporal direction with respect to the local time QRF $t_A$ or $t_B$.
\end{itemize}
\noindent
The first of these points enforces the informational symmetry of $\mathscr{B}$ and hence enforces unitarity.  The second requires any ``carrier'' of information to be its own antiparticle.  

Following this idea of information as subject to scattering and representing the total initial and final informational states of $k = ~A$ or $B$ as $|out \rangle^k$ and $|in \rangle^k$ respectively, we can write an $S$-matrix:

\begin{equation}
S^k : |out \rangle^k \mapsto |in \rangle^k\,,
\end{equation}
\noindent
where the reversal of the usual order reflects the fact that information always flows {\em from} a preparation device {\em to} a measurement device.  This $S^k$ is $N \times N$; Figs. \ref{qubit-screen-fig} and \ref{scattering-fig} depict it in bases in which it is diagonal.  As $S^k$ is simply a representation of Eq. \eqref{ham} for system $k$, we have the expected conclusion:

\begin{quote}
Every interaction between separable systems can be represented as scattering.
\end{quote}
\noindent
This conclusion rests solely on the GHP, requiring no assumptions about an embedding geometry.

Adding the geometry shown in Fig.~\ref{qubit-fig} to $\mathscr{B}$ allows $\mathscr{B}$ to be viewed as a ``physical'' horizon, e.g. the (stretched) horizon of a BH, the motivating system of interest for both the Bekenstein area law and the HP.  Figure \ref{scattering-fig} then represents traversal of or scattering from the horizon as observed by either the BH interior $B$ or the external ``rest of the universe'' $A$.  Coupled pair production events near the horizon yield symmetric diagrams of this kind \cite{fgm:21}; hence Fig.~\ref{scattering-fig} is consistent with the observation of Hawking radiation by an asymptotic observer ``embedded'' in $A$.  That the formation and evaporation of a BH could be considered a scattering process was originally proposed by 't Hooft \cite{tHooft:96} and has since been given an explicit formulation \cite{betzios:16}.

It is important to emphasize that the GHP requires and Fig. \ref{scattering-fig} represents a fixed decomposition $U = AB$ for which $|AB \rangle = ~|A \rangle |B \rangle$ and hence fixed Hilbert-space dimensions for $A$ and $B$.  It does not, therefore, represent net transfers of degrees of freedom across $\mathscr{B}$.  ``Collapse'' or ``infall'' processes and, dually, ``evaporation'' processes alter the interior geometry of $B$ but not its Hilbert-space dimension.  The information encoded on $\mathscr{B}$ can be altered by these processes, but the dimension $N$ and the topology of $\mathscr{B}$ remain invariant.  Such topologically-invariant models of BH evolution have been developed previously based on the proposed implementation of entanglement by Einstein-Rosen bridges \cite{susskind:13, susskind:14}.

\section{HP-like principles appear in multiple disciplines} \label{history}

We have thus far considered the physical meaning of the HP, as generalized to the GHP, from the perspective of quantum information theory.  We now broaden this perspective to consider principles analogous to the HP that have been formulated independently in other disciplines.  Such principles have received wide-spread application, suggesting that the HP is in fact a general principle of not just of quantum theory, but of all of science.

\subsection{Markov blankets and the FEP}\label{mb-1}

\begin{figure}[H]
\centering
\includegraphics[width=12 cm]{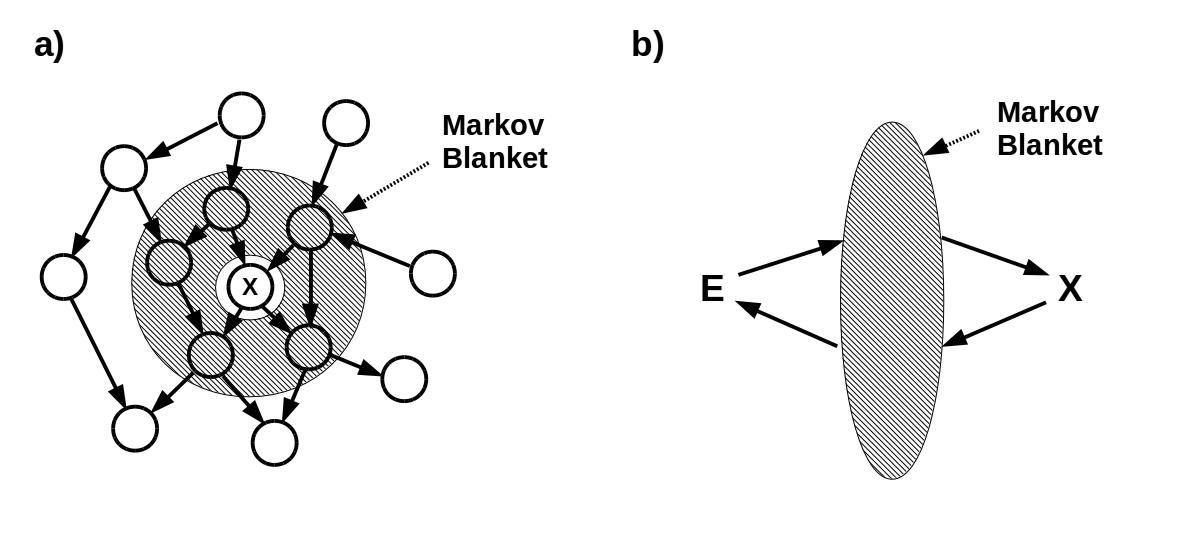}
\caption{a) The MB of a node X in a causal network comprises the parents and children of X together with any other parents of X's children. b) The MB is effectively an information channel separating X from its environment E.  From \cite{fl:17}, Fig. 2; used with permission.}
\label{MB-fig}
\end{figure}

The idea of a Markov blanket (MB) was formulated by Pearl \cite{pearl:88} to capture the emergent conditional independence of disjoint components of finite causal networks with Markovian dynamics by judiciously defining the boundaries of the systems in question.  The MB of any node or subnetwork X of such a causal network comprises all nodes that are parents of X (i.e. nodes with arrows to X), children of X (i.e. nodes with arrows from X), or other parent's of X's children, as shown in Fig. \ref{MB-fig}.  All information exchanged between the node or subnetwork X and the nodes exterior to its MB must traverse the MB; the MB thus functions as a finite classical information channel between X and its external ``environment'' E.  The separation between X and E imposed by the MB renders them mutually conditionally independent\footnote{In many applications, the MB is partitioned into sensory and active states mediating the influence of external upon internal states, and vice-versa. Specifically, external states act on sensory states which influence, but are not influenced by, internal states. Internal states couple back via active states which influence but are not influenced by external states \cite[Table 1]{palacios:20}.}.

Friston \cite{friston:19} has shown that any random dynamical system that has a non-equilibrium steady state (NESS) solution to its density dynamics i) has an internal dynamics that is conditionally independent of the dynamics of its environment, and hence has a MB, and ii) will continuously ``self-evidence'' by returning its state to (the vicinity of) its NESS.  Satisfying these conditions is required of any system that is observable as such over time, i.e. any system for which sequential measurements as considered in \S \ref{TQFTs} above.  Such systems can be described as minimizing a variational free energy (VFE) functional that effectively measures their uncertainty about their environment's future behavior.  The free energy principle (FEP) is the statement that any random dynamical system meeting the above two conditions, i.e. any system for which sequential measurements are possible, will behave in a way that asymptotically minimizes its detected VFE.  This way of characterizing random dynamical systems gives rise to a ``Bayesian mechanics'' \cite{ramstead:22} that reformulates classical physics in decision-theoretic language within a scale-free computational architecture that is applicable, in principle, from the molecular and cellular levels up to the cosmological. In such a model, aptly described as Bayesian selection, natural selection itself can be viewed as structure learning based upon the model evidence encoded by some phenotype \cite{palacios:20}.

Using the tools reviewed in \S \ref{theory-meas} above, we have shown \cite{ffgl:22} that the FEP can be reformulated for generic, finite quantum systems meeting the separability required by Eq. \eqref{ham}.  The MB is, in this case, implemented by a holographic screen $\mathscr{B}$ compliant with the GHP.  Strict minimization of VFE drives systems to share QRFs across $\mathscr{B}$, i.e. drives them to entanglement as discussed in \S \ref{entanglement} above.  The FEP is, therefore, asymptotically a restatement of the Axiom of Unitarity.

\subsection{Multiple realizability and virtual machines}

The foundational principle of computer science is the Church-Turing (CT) thesis, which states that any computable function can be computed by the $\lambda$-calculus \cite{church:36} or a Turing machine \cite{turing:36}.  While the CT is often regarded simply as establishing two universal models of computation, at a deeper level it states the multiple realizability of computation: any process that emulates, or can be emulated by, the $\lambda$-calculus or a Turing machine can be considered a computational process.  The CT thesis thus underlies a definition of computation in terms of emulation: any process that can be (usefully) interpreted as a computation is a computation \cite{horsman:14}.

As emphasized in \cite{horsman:14}, interpreting a physical process as a computation relies on finite-resolution observations of some finite number of sequential states.  A process useful as a computer must, moreover, allow manipulations that return it to some (quasi-) stable state from which it can be perturbed into a set of distinct ``input'' states.  Whether an arbitrary such system will reach a (quasi-) stable state and hence ``halt'' after a finite number of computational steps from some finite input cannot be determined algorithmically \cite{turing:36}; this is the Halting problem \cite{hopcroft:79}.  Whether a finite, step-by-step description of the observed behavior of an arbitrary system following any one or more of some circumscribed set of input perturbations specifies a computation of some nontrivial function is similarly algorithmically undecidable (Rice's theorem \cite{rice:53}).

The concept of a ``black box'' (BB) formulated in classical cybernetics provides an alternative statement of multiple realizability that does not depend explicitly on the theory of computation \cite{ashby:56}.  A BB is similarly a physical system that permits finite numbers of finite-resolution perturbations and observations.  The interior of the BB is considered to contain a ``machine table'' that determines the next output given the history of inputs.  Finite sequences of perturbations and observations are insufficient to determine the machine table of an arbitrary BB (Moore's theorem \cite{moore:56}).

In practice, multiple realizability allows multiple distinct hardware architectures to compute the same functions, and
multiple programming languages with widely differing syntax and semantics to have the same computational power.  It enables layered computing architectures in which each layer treats the layers both above and below as virtual machines (VMs) that have specified, finite application programming interfaces (APIs) but are otherwise unconstrained in implementation \cite{smith:05}.  The top-level VM is the user interface, which allows the user finite manipulations and observations of the behavior of the underlying architecture while ``hiding'' implementation details.  While the implementation details of practical computers are accessible in principle, reverse-engineering them from behavioral observations becomes increasingly difficult as the depth of architectural layering increases, and rapidly becomes intractable if components are distributed across a network that supports asynchronous communication.

As quantum systems, quantum computers encode nonfungible information, rendering the reverse engineering problem unsolvable in principle.  Quantum artificial neural networks (QNNs) generalize classical artificial neural networks, which are Turing equivalent.  Conventional QNNs can be further generalized to {\em topological QNNs} (TQNNs), which as structured as spin networks, are tensor network representations of TQFTs, and hence fully compliant with the GHP as discussed in \S \ref{TQFTs} above \cite{fgm:22, marciano:22}. It is in view of such tensor networks, and the development of several sections here in relation to the boundary $\mathscr{B}$ (as in e.g
\S \ref{entanglement}) that further open up connections with the AdS/CFT correspondence to be pursued in future work. Enticing is the question: ``is spacetime a quantum error-correcting code (QECC)?'' (reviewed in \cite{bain:20}). Different slants and interpretations of this question are discussed in \cite{bain:20}. For instance, the hypotheses of
\cite{almheiri:15,pastawski:15}, involve space in the bulk as emerging from boundary systems that can realize the structure of a QECC;  \cite{vanraamsdonk:11} suggests that the connectivity of spacetime in the bulk is related to the entanglement structure of the codespace of a boundary CFT subject to its QECC.

\subsection{Active inference and interface theories}

The primary application of the FEP has been to biological systems, where it underpins the idea of Bayesian ``active inference'' in which living systems increase their predictive power not just through learning, but also through active manipulations of their environments \cite{friston:10, friston:13, ramstead:18}.  This principle of active inference, or curiosity-driven learning, is both scale-independent and applicable not just to motions or other actions in 3d space but also to actions in more abstract state spaces, e.g. those of the genome or the metabolic system \cite{fl:20, fl:22}.  Indeed we have recently shown that the FEP drives the high fan-in, high fan-out ``neuromorphic'' organization of sensory and effector systems that is ubiquitous across biology at all scales \cite{ffglm:22}.

The goal of any system employing active inference is to reduce VFE over the long term by learning to predict how its environment will act on its MB.  Crucially, enacting self-organization necessitates the emergence of boundaries defining the separation of internal from external states \cite{palacios:20} (and \S\ref{mb-1} here). Prediction is accomplished by a computational system that is a generative model, in the sense of the Good Regulator Theorem \cite{conant:70}, of its environment as represented on its MB.  Such models may incorporate ``metacognitive'' components that represent the system itself, again via the system's actions on its MB, which include memory writes as discussed in \S \ref{systems} above \cite{kfl:22}.  Active inference systems have, by definition, no direct access to their environments beyond their MBs.  The MB acts, in this case, as a system-environment interface, in the sense of an API.

Interface theories of perception and action have also been developed independently of the FEP, particularly by Hoffman and colleagues \cite{hoffman:09, hoffman:12, hoffman:15}, who also show explicitly that natural selection processes do not favor ``veridical'' perception beyond the interface \cite{hoffman:13, prakash:20, singh:21}.  Both spacetime and perceived ``objects'' are explicitly emergent from computational processes implemented by the perceiving agent -- effectively, its generative model -- in this theoretical setting \cite{fhpp:17}.

\section{Conclusion}

We have shown here that the HP, particularly when generalized to the GHP, is not merely ``an apparent law of physics that stands by itself'' but rather a deep, foundational principle.  It is a principle of restricted access.  It lays a non-negotiable price on separability: if two systems are separated by a boundary $\mathscr{B}$, their access to each other is limited to the information $\mathscr{B}$ itself can encode.  When stated in this way, the HP seems shockingly obvious.  When its implications for our ordinary concepts of ``objects'' and ``spacetime'' are pointed out, however, it can seem deeply mysterious.  The idea that classical information is decomposition-relative -- and hence is {\em observer}-relative -- strongly challenges our pretheoretical sense of an ``objective reality'' shared by all physical systems.  As Wheeler \cite{wheeler:83} points out, this challenge lies at the very heart of quantum theory.

What is perhaps most significant about the HP, however, is its emergence over the past century as a foundational principle not just of physics, but of all disciplines that directly address information transfer between separated systems.  Its ubiquity speaks simultaneously to the fundamental unity of science and to its fundamental limitations as an empirical enterprise.

\section*{Acknowledgments}
C.F. acknowledges support from the John Templeton Foundation (Grant No. TBA).  A.M.~wishes to acknowledge support by the Shanghai Municipality, through the grant No.~KBH1512299, by Fudan University, through the grant No.~JJH1512105, the Natural Science Foundation of China, through the grant No.~11875113, and by the Department of Physics at Fudan University, through the grant No.~IDH1512092/001.


\begin{thebibliography}{}

\bibitem{tHooft:93}
't Hooft, G. Dimensional reduction in quantum gravity. In Ali, A., Ellis, J., Randjbar-Daemi, S. (eds) Salamfestschrift. Singapore: World Scientific, 1993, pp. 284--296.

\bibitem{bekenstein:73}
Bekenstein, J. D.  Black holes and entropy.  Phys. Rev. D 1973;{7}:2333--2346.

\bibitem{susskind:95}
Susskind, L. The world as a hologram. J. Math. Phys. 1995;{36}:6377--6396.

\bibitem{bousso:02}
Bousso, R. The holographic principle. Rev. Mod. Phys. 2002; \textbf{74}:825--874.

\bibitem{rovelli:17}
Rovelli, C. Black holes have more states than those giving the Bekenstein-Hawking entropy: A simple argument. Preprint arXiv:1710:00218.

\bibitem{rovelli:19}
Rovelli, C. The subtle unphysical hypothesis of the firewall theorem. Entropy 2019;21:839.

\bibitem{almheiri:21}
Almheiri, A., Hartman, T., Maldacena, J., Shaghoulian, E., Tajdini, A. The entropy of Hawking radiation.  Rev. Mod. Phys. 2021;93:035002.

\bibitem{maldecena:98}
Maldecena, J. The large N limit of superconformal field theories and supergravity. Adv. Theor. Math. Phys. 1998;2:231--252. (1998)

\bibitem{strominger:01}
Strominger, A. The dS/CFT correspondence.  J. High Energy Phys. 2001;10:034.

\bibitem{fm:19}
Fields, C., Marcian\`{o}, A.  Sharing nonfungible information requires shared nonfungible information. Quant. Rep. 2019;1: 252--259.

\bibitem{fm:20}
Fields, C., Marcian\`{o}, A. Holographic screens are classical information channels.  Quant. Rep. 2019;2;326--336.

\bibitem{fg:20}
Fields C, Glazebrook JF. Representing measurement as a thermodynamic symmetry breaking. Symmetry 2020;12:810.

\bibitem{fgm:21}
Fields C, Glazebrook JF, Marcian\`{o} A.  Reference frame induced symmetry breaking on holographic screens. Symmetry 2021;13:408.

\bibitem{addazi:21}
Addazi A, Chen P, Fabrocini F, Fields C, Greco E, Lulli M, Marcian\`{o} A, Pasechnik R. Generalized holographic principle, gauge invariance and the emergence of gravity \`{a} la Wilczek. Front Astron Space Sci 2021;8:563450.

\bibitem{macclane:94}
Mac Clane, S., Moerdijk, I. Sheaves in Geometry and Logic: A First Introduction to Topos Theory. Universitext Series. Springer-Verlag, New York-Heidelberg, 1994.

\bibitem{fgm:22}
Fields C, Glazebrook JF, Marcian\`{o} A.  Sequential measurements, topological quantum field theories, and topological quantum neural networks.  Fortschr. Phys. 2022;2200104, 26 pp.,in press.

\bibitem{ffgl:22}
Fields, C., Friston K., Glazebrook J.F., Levin M.  A free energy principle for generic quantum systems.  Prog. Biophys. Mol. Biol. 2022;173:36--59.

\bibitem{friston:10}
Friston, K. J. The free-energy principle: A unified brain theory? Nat. Rev. Neurosci. 2010;11:127--138.

\bibitem{friston:13}
Friston, K. J. Life as we know it.  J. R. Soc. Interface 2013;10:20130475.

\bibitem{ramstead:18}
Ramstead, M. J., Badcock, P. B., Friston, K. J. Answering Schr\"{o}dinger’s question: a free-energy formulation. Phys. Life Rev. 2018;24:1--16.

\bibitem{friston:19}
Friston, K. J. A free energy principle for a particular physics.  Preprint arxiv:1906.10184 [q-bio.NC].

\bibitem{ramstead:22}
Ramstead, M. J., Sakthivadivel, D. A. R., Heins, C., Koudahl, M., Millidge, B., Da Costa, L., Klein, B., Friston, K. J. On Bayesian mechanics: A physics of and by beliefs.  Preprint arxiv:2205.11543 [cond-mat.stat-mech].

\bibitem{pegg:02}
Pegg, D., Barnett, S., Jeffers, J. Quantum theory of preparation and measurement.  J. Mod. Optics 2002;49:913--924.

\bibitem{nielsen:00}
Nielsen MA, Chuang IL. Quantum Computation and Quantum Information.  New York: Cambridge University Press, 2000.

\bibitem{vonN:55}
Von Neumann J. The Mathematical Foundations of Quantum Mechanics. Princeton University Press, Princeton, NJ, 1955.

\bibitem{landauer:61}
Landauer R. Irreversibility and heat generation in the computing process. IBM J Res Devel 1961;5:183--195.

\bibitem{landauer:99}
Landauer R. Information is a physical entity. Physica A 1999;263:63--67.

\bibitem{bennett:82}
Bennett CH. The thermodynamics of computation. Int J Theor Phys 1982;21:905--940.

\bibitem{everett:57}
Everett, H. III ``Relative state'' formulation of quantum mechanics.  Rev. Mod. Phys. 1957;29:454--462.

\bibitem{tegmark:98}
Tegmark, M. The interpretation of quantum mechanics: Many worlds or many words? Fortschr. Phys. 1998;46:855--862.

\bibitem{rovelli:96}
Rovelli, C. Relational quantum mechanics. Int. J. Theor. Phys. 1996;35:1637--1678.

\bibitem{fuchs:10}
Fuchs C. QBism, the Perimeter of quantum Bayesianism. Preprint arxiv:1003.5209, 2010.

\bibitem{mermin:18}
Mermin ND. Making better sense of quantum mechanics. Rep Prog Phys 2018;82:012002.

\bibitem{bohr:34}
Bohr N. Atomic Theory and the Description of Nature.  Cambridge University Press, Cambridge, UK, 1934.

\bibitem{mermin:98}
Mermin ND. What is quantum mechanics trying to tell us? Am. J. Phys. 1998;66:753--767.

\bibitem{schrodinger:35}
Schr\"{o}dinger, E. Die gegenw\"{a}rtige Situation in der Quantenmechanik. Naturwissenschaften 1935;23:807--812,823--828,844--849.  Translated in: Wheeler, J. A., Zurek, W. H. Quantum Theory and Measurement.  Princeton Unversity Press, Princeton, NJ, 1983, pp. 152--167.

\bibitem{landsman:07}
Landsman, N. P.  Between classical and quantum. In: Butterfield J, Earman J. (eds.), Handbook of the Philosophy of Science: Philosophy of Physics. Elsevier, Amsterdam, Netherlands, 2007, pp. 417--553.

\bibitem{cabello:15}
Cabello, A. Interpretations of quantum theory: A map of madness.  In: Lombardi, O. et al. (eds), What is Quantum Information? Cambridge Unversity Press, Cambridge, UK, 2017, pp. 138--143.

\bibitem{wheeler:89}
Wheeler JA. Information, physics, quantum: The search for links. In Zurek W (ed) Complexity, Entropy, and the Physics of Information. Boca Raton, FL: CRC Press, 1989; pp. 3--28.

\bibitem{aharonov:84}
Aharonov Y, Kaufherr T. Quantum frames of reference. Phys Rev D 1984;30:368--385.

\bibitem{bartlett:07}
Bartlett SD, Rudolph T, Spekkens RW. Reference frames, super-selection rules, and quantum information. Rev Mod Phys 2007;79:555--609.

\bibitem{conway:09}
Conway JH, Kochen S.  The strong free will theorem.  Notices AMS 2009;56(2):226--232.

\bibitem{tipler:14}
Tipler, F. Quantum nonlocality does not exist.  Proc. Natl. Acad. Sci. USA 2014;111:11281--11286.

\bibitem{barwise:97}
Barwise J, Seligman J, Information Flow: The Logic of Distributed Systems (Cambridge Tracts in Theoretical Computer Science, 44). Cambridge, UK: Cambridge University Press, 1997.

\bibitem{fg:19a}
Fields C, Glazebrook JF. A mosaic of Chu spaces and Channel Theory I: Category-theoretic concepts and tools.  J Expt Theor Artif Intell 2019;31:177--213.

\bibitem{barr:79}
Barr, M. (1979). *-Autonomous categories, with an appendix by Po Hsiang Chu. \emph{Lecture Notes in Mathematics } {752} (1979). Berlin: Springer.

\bibitem{pratt:97}
Pratt, V. Types as Processes, via Chu spaces. Invited paper. Proceedings `Express'97: Fourth Workshop on Expressiveness in Concurrency'.
Santa Margherita, Italy. September 1997. \emph{Electronic Notes in Theoretical Computer Science} {7} 21pp.

\bibitem{pratt:99a}
Pratt, V. Chu spaces. \emph{School on Category Theory and Applications (Coimbra 1999)}, Vol. 21 of \emph{Textos Mat.
S\'{e}r. B}, University of Coimbra, Coimbra, 1999 (pp. 39--100).

\bibitem{pratt:00}
Pratt, V. Higher dimensional automata revisited. \emph{Mathematical Structures in Computer Science} \textbf{10}(4) (2000), 525--548.

\bibitem{seigal:22}
Seigal, A., Harrington, H. A., and Nanda, V., Principal components along quiver representations.
{\em Found Comput Math} (2022)
https://doi.org/10.1007/s10228-022-09563-x
%%arXiv:2104.10666v2[math.RT]

\bibitem{fg:22}
Fields C, Glazebrook JF. Information flow in context-dependent hierarchical Bayesian inference. J Expt Theor Artif Intell 2022;34:111--142.

\bibitem{abramsky:11}
Abramsky, S. and  Brandenburger, A. The sheaf-theoretic structure of non-locality and contextuality. \emph{New Journal of Physics} {13} (2011), 113036

\bibitem{dzha:17b}
Dzhafarov, E. N. Cervantes, V. H. and Kujala, J. V. Contextuality in  canonical systems of random variables.
{Philosophical Transactions of The Royal Society A} {375} (2017), 20160389
%%doi:10.1098/rsta.2016.0389

\bibitem{gudder:19}
Gudder, S. Contexts in quantum measurement theory. {Foundations of Physics} {49} (2019), 647--662.

\bibitem{mermin:90}
Mermin, N. D. Simple unified form for the major no-hidden variables theorem. {\em Phys. Rev. Lett.} {65} (1990), 3373.

\bibitem{howard:14}
Howard, M., Wallman, J., Veitch, V. and Emerson, J. Contextuality supplies the `magic' for quantum computation. {\em Nature} {510} (2014), 351--353.

\bibitem{bermejo:17}
Bermejo-Vega, J., Delfosse, N. Browne, D. E., Okay, C. and Raussendorf, R. Contextuality as a resource for models of quantum computation with qubits.
{\em Phys. Rev. Lett.} {119} (2017), 120505.


\bibitem{marciano:22}
Marcian\`o A., Chen D., Fabrocini F., Fields C., Greco E., Gresnigt N., Jinklub K., Lulli M., Terzidis K. and Zappala E. Quantum neural networks and topological quantum field theories. Neural Networks 2022;153:164--178.

\bibitem{atiyah:88}
Atiyah, M.  Topological quantum field theory. Publ. Math. IH\'{E}S 1988;68:175--186.

\bibitem{quinn:95}
Quinn, F. (1995). Lectures on axiomatic topological quantum field theory. In {\em Geometry and quantum field theory}. Lecture notes from the graduate summer school program, June 22-July 20, Park City UT, USA, pp. 323--453. {\em Amer. Math. Soc.}, Providence RI. Instiitute for Advanced Studies, Princeton NJ.

\bibitem{shannon:48}
Shannon, C. E. A mathematical theory of communication. Bell Syst. Tech. J. 1948;27:379--423

\bibitem{tegmark:12}
Tegmark M. How unitary cosmology generalizes thermodynamics and solves the inflationary entropy problem. Phys. Rev. D 2012;85:123517.

\bibitem{rovelli:21}
Di Biagio A, Don\`{a} P, Rovelli C. The arrow of time in operational formulations of quantum theory.  Quantum 2021;5;520.

\bibitem{baez:20}
Baez JC.  Getting to the bottom of Noether's theorem.  Preprint arxiv:2006.14741, 2020.

\bibitem{ryu:06}
Ryu, S. and Takayanagi, T. Holographic derivation of entanglement entropy from AdS/CFT. Phys. Rev. Lett. {96} (2006), 181602.

\bibitem{headrick:14}
Headrick, M., Hubeny, V. E., Lawrence, A. and Rangamani, M. Causality \& holographic entanglement entropy. {\em Journal of High Energy Phyics} {12}(214), 162.  (2014), 133.

\bibitem{casini:04}
Casini, H. Geometric entropy, area and strong subadditivity. {\em Class. Quant. Grav.} {21} (2004), 2351.

\bibitem{lee:11}
Lee, J.-W. Quantum entanglement from the holographic principle. Preprint arXiv:1109.3542v1 [hep-th].

\bibitem{Qi:2013caa}
Qi, X.~L. Exact holographic mapping and emergent space-time geometry,
arXiv:1309.6282 [hep-th].

\bibitem{Han:2016xmb}
Han M.  Hung, L.~Y. Loop Quantum Gravity, Exact Holographic Mapping, and Holographic Entanglement Entropy, Phys. Rev. D \textbf{95} (2017) no.2, 024011.
%doi:10.1103/PhysRevD.95.024011
%arXiv:1610.02134 [hep-th].

\bibitem{Mertens:2017mtv}
Mertens, T.~G. Turiaci, G.~J. Verlinde, H.~L. Solving the Schwarzian via the Conformal Bootstrap, JHEP \textbf{08} (2017), 136.
%doi:10.1007/JHEP08(2017)136
%arXiv:1705.08408 [hep-th].

\bibitem{Blommaert:2018oro}
Blommaert, A. Mertens, T.~G. Verschelde, H. The Schwarzian Theory - A Wilson Line Perspective, JHEP \textbf{12} (2018), 022.
%doi:10.1007/JHEP12(2018)022
%[arXiv:1806.07765 [hep-th]].

\bibitem{wootters:01}
Wootters, K. Entanglement of formation and concurrence. Quant. Inform. Comp. 2001;1(1):27--44.

\bibitem{wooters:82}
Wootters WK, Zurek WH. A single quantum cannot be cloned.  Nature 1982;299:802--803.

\bibitem{tHooft:96}
't Hooft, G. The scattering matrix approach for the quantum black hole: An overview. Int. J. Mod. Phys. A 1996;11:4623--4688.

\bibitem{betzios:16}
Betzios, P., Gaddam, N., Papadoulaki, O.  The black hole S-Matrix from quantum mechanics.  J. High Energy Phys. 2016;11:131.

\bibitem{susskind:13}
Maldacena, J., Susskind, L. Cool horizons for entangled black holes. Fortschr. Phys. 2013;61:781--811.

\bibitem{susskind:14}
Susskind, L. Entanglement is not enough. Preprint arXiv:1411.0690, 2014.

\bibitem{pearl:88}
Pearl J. Probabilistic Reasoning in Intelligent Systems: Networks of Plausible Inference. Morgan Kaufmann, San Mateo, CA, 1988.

\bibitem{palacios:20}
Palacios, E. R., Razi, A., Parr, T., Kirchoff, M. and Friston, K. On Markov blankets and hierarchical self-organization. \emph{J. Theoretical Biology} {486} (2020), 110089.

\bibitem{fl:17}
Fields, C., Levin, M. Multiscale memory and bioelectric error correction in the cytoplasm-cytoskeleton-membrane system.  WIRES Syst. Biol. Med. 2018;10:e1410.

\bibitem{church:36}
Church, A. A Note on the {\em Entscheidungsproblem}. J. Symbol. Logic 1936;1:40--41.

\bibitem{turing:36}
Turing, A. M. On computable numbers, with an application to the {\em Entscheidungsproblem}. Proc. London Math.
Soc. Ser. 2 1936;42:230--265.

\bibitem{horsman:14}
Horsman C, Stepney S, Wagner RC, Kendon V. When does a physical system compute? Proc R Soc A 2014;470:20140182.

\bibitem{hopcroft:79}
Hopcroft, J. E., Ullman, J. D.  Introduction to Automata Theory, Languages, and Computation.  Boston, Addision-Wesley, 1979.

\bibitem{rice:53}
Rice, H. G. Classes of recursively enumerable sets and their decision problems. Trans. Am. Math. Soc. 1953;74:358--366.

\bibitem{ashby:56}
Ashby, W.R. Introduction to Cybernetics. London: Chapman and Hall, 1956.

\bibitem{moore:56}
Moore, E. F. Gedanken experiments on sequential machines. In: Shannon, C. W.; McCarthy, J. (Eds.) Autonoma Studies. Princeton, NJ: Princeton University Press, 1956, pp. 129–155.

\bibitem{smith:05}
Smith, J. E., Nair, R. The architecture of virtual machines.  IEEE Computer 2005;38(5):32--38.

\bibitem{bain:20}
Bain, J. Spacetime as a quantum error correcting code? {\em Studies in History and Philosophy of Science Part B: Studies in History and Philosophy of Modern Physics} {71} (2020), 26--36.

\bibitem{almheiri:15}
Almheiri, A. Dong. X. and Harlow, D. Bulk locality and quantum error correction in AdS/CFT. {\em Journal of High Energy Phyics} {4}(163) (2015), 133.

\bibitem{pastawski:15}
Pastawski, F., Yoshida, B., Harlow, D. and Preskill, J. Holographic quantum error-correcting codes: toy models for the bulk/boundary correspondence.
{\em Journal of High Energy Physics} {6}(149) (2015), 53 pp.
doi:10.1007/JHEP06(2015)149

\bibitem{vanraamsdonk:11}
Van Raamsdonk, M. Building up spacetime with quantum entanglement. {\em General Relativity and Gravitation} {42} (2011), 23232329.


\bibitem{fl:20}
Fields, C., Levin, M. Integrating evolutionary and developmental thinking into a scale-free biology. BioEssays 2020;42:1900228.

\bibitem{fl:22}
Fields, C., Levin, M. Competency in navigating arbitrary spaces as an invariant for analyzing cognition in diverse embodiments. Entropy 2022;24:819.

\bibitem{ffglm:22}
Fields C, Friston, K., Glazebrook JF, Levin, M., Marcian\`{o} A. The Free Energy Principle drives neuromorphic development.  Preprint arxiv:2207.09734, 2022.

\bibitem{conant:70}
Conant RC, Ashby WR.  Every good regulator of a system must be a model of that system.  Int. J. Syst. Sci. 1970;1(2):89--97.

\bibitem{kfl:22}
Kuchling F, Fields C, Levin M. Metacognition as a consequence of competing evolutionary time scales. Entropy 2022;24:601.

\bibitem{hoffman:09}
Hoffman, D. D. The interface theory of perception. In: Dickinson, S., Tarr, M., Leonardis, A., Schiele, B. (Eds.) Object Categorization: Computer and Human Vision Perspectives, Cambridge University Press: New York,NY, 2009, pp. 148--165.

\bibitem{hoffman:12}
Hoffman, D. D., Singh, M. Computational evolutionary perception. Perception 2012;41:1073--1091.

\bibitem{hoffman:15}
Hoffman, D. D., Singh, M., Prakash, C. The interface theory of perception.  Psychon. Bull. Rev. 2015;22:1480--1506

\bibitem{hoffman:13}
Hoffman D. D., Singh, M., Mark, J. T. Does natural selection favor true perceptions? Human Vision Electron. Imaging XVIII (SPIE 18), 2013, 865104.

\bibitem{prakash:20}
Prakash C, Fields C, Hoffman DD, Prentner R, Singh M.  Fact, fiction, and fitness. Entropy 2020;22:514.

\bibitem{singh:21}
Prakash C, Stephens KD, Hoffman DD, Prentner R, Singh M, Fields C. Fitness beats truth in the evolution of perception.  Acta Biotheor. 2021;69:319--341.

\bibitem{fhpp:17}
Fields, C., Hoffman, D. D., Prakash, C., Prentner, R. Eigenforms, interfaces, and holographic encoding.  Construct. Found. 2017;12:265--291.

\bibitem{wheeler:83}
Wheeler, J.A. Law without law. In Quantum Theory and Measurement; Wheeler, J.A., Zurek, W.H., Eds.;
Princeton University Press: Princeton, NJ, USA, 1983; pp. 182--213.


\end{thebibliography}
\end{document}